\documentclass[onecolumn, 11pt, draftcls]{IEEEtran}
\def\app#1{
\addtocounter{section}{1}
\setcounter{equation}{0}
\renewcommand{\thesection}{\Alph{section}}
\renewcommand{\theequation}{\Alph{section}.\arabic{equation}}
\vspace{7mm}
\begin{center}{\normalsize A}{\footnotesize PPENDIX} {\normalsize \Alph{section}\\
{\footnotesize #1}}
\end{center}
\vspace{5mm}
}

\newcommand{\TS}{

\vspace{-1cm}
\noindent
}

\usepackage[dvipdfmx]{graphicx}

\usepackage{latexsym}
\usepackage{amsmath}

\newcommand{\LLL}{\limsup_{n\to\infty}\frac{1}{n}\log}
\newcommand{\lll}{\liminf_{n\to\infty}\frac{1}{n}\log}

\newcommand{\Eq}{\begin{equation}}
\newcommand{\Qe}{\end{equation}}
\newcommand{\Eqa}{\begin{eqnarray}}
\newcommand{\Qea}{\end{eqnarray}}
\newcommand{\Eqas}{\begin{eqnarray*}}
\newcommand{\Qeas}{\end{eqnarray*}}
\newcommand{\No}{\nonumber\\}
\def\C#1{{\cal #1}}
\def\B#1{{\bf #1}}
\def\T#1{\widetilde{#1}}
\def\H#1{\widehat{#1}}

\def\O#1{\overline{#1}}

\newcommand{\Ve}{\varepsilon} 

\newcommand{\X}{\B{X}}

\newcommand{\LSUPn}{\limsup_{n\rightarrow\infty}}
\newcommand{\LINFn}{\liminf_{n\rightarrow\infty}}

\newcommand{\Tlad}{\vspace{2mm}}
\newcommand{\Prend}{\hspace*{\fill}$\Box$}
\newcommand{\SP}{\vspace{10mm}}

\newtheorem{theorem}{\SP \it Theorem\rm}
\newtheorem{lemma}{\SP \it Lemma\rm}

\newtheorem{define}{\SP \it Definition\rm}
\newtheorem{prop}{\SP \it Proposition\rm}

\newtheorem{remark}{\SP \it Remark\rm}

\renewcommand{\theequation}{\arabic{section}.\arabic{equation}}

\renewcommand{\thesection}{\Roman{section}}

\title{
A Method of Expressing the Magnitude of Merit of Being Able to Access a Side Information During Encoding
}
\author{
Kiminori Iriyama
\thanks{This paper was presented at the Symposium on Information Theory and its Applications (SITA), Fukushima, Japan, 
December 2018.}
}

\begin{document}
\maketitle

\abstract{

In general, if there is one device $A$ with the same performance as many devices $B$, it would be better to replace many devices with one device. 
In order to determine the number of devices that can be reduced, it is important to determine the number of other devices $B$ having the same performance as one device $A$.
In this paper, based on the concept of ``representing the performance of one coding system A\_n by the number of other coding systems B\_n", 
we consider the merits of an encoder that can access the side information in the fixed-length source coding in general sources. 
The purpose of this paper is to characterize the merit of being able to access the side information during encoding with the number k\_n of coding systems that can not access the side information during encoding.
For this purpose, we derive general formulas for the upper limit of error probability and reliability functions without assuming a specific information theoretical structure.
These general formulas are applicable to various fields other than source coding problems, because they have been proved without assuming any particular information theoretical structure.
We prove some theorems under the condition that the coding system is subadditive.
}


\section{
Introduction}
We are interested in the fixed-length source coding with side information. 
In this paper, a code that can access side information in both encoding and decoding is called an $A_n$-code, and a code that can access the side information only during decoding is called a $B_n$-code. 
The performance of the encoder is evaluated by the coding rate and the error probability.
Among the encoders having the same coding rate, it can be said that the performance is better as the encoder having a smaller error probability.
If the optimal $A_n$-code has a performance corresponding to $k_n+1$ $B_n$-codes, the magnitude of the merit of accessing side information during encoding may be considered to correspond to  $k_n$ $B_n$-codes.
This reason is described below. 
First, let us assume a situation where the side information can not be accessed during encoding. 
Next, suppose we select the following (a) or (b) to improve the encoder performance: 
\begin{enumerate}
\item[(a)] Allows access to the side information during encoding. 
\item[(b)] Add $ k_n $ $ B_n $-codes. 
\end{enumerate}
Option (a) corresponds to choosing an $A_n$-code. 
Option (b) corresponds to selecting $k_n+1$ $B_n$-codes. 
If the performance of the optimal $A_n$-code corresponds to that of $k_n + 1 $ $ B_n $-codes, then options (a) and (b)  may be considered to have equivalent performance.
That is, the merit of being able to access the side information during encoding may be considered to be equal to the merit of adding $k_n$ $B_n$-codes. 
Our aim is to characterize the number $k_n$ under the condition that the coding rate is smaller than the given $R >0$. 
One of the main results of this paper is that,  in the high-rate case,
$k_n= [n^s]$  ($ s> 1 $) (Theorem \ref {th:ABr}). 
When not at high coding rates, 
$k_n=0$ (Theorem \ref{th:ABe}, \ref{th:ABr*}).

In order to prove theorems about $k_n$, we give some theorems about minimum achievable coding rates and reliability functions without assuming any particular information theoretical structure.
In particular, we establish a general formula for the reliability function $\rho_C(R|\mu ) $ in the high-rate case and is one of the main results of this paper (Theorem \ref{th:Cr}).
The concept of "subadditive" introduced in this paper plays an important role in the proof of Theorem  \ref{th:Cr}. 
The upper limit $\Ve_C(R|\mu )$ of the error probability  $\Ve_n$ of the optimum code is expressed in terms of $G_u(\nu \| \mu)$ which is introduced in this paper.
The formula of $\Ve_C(R|\mu )$ is interesting because it can be rewritten as the formula for the reliability function $\rho^*(R|\mu)$  in the low-rate case by replacing  $G_u(\nu \| \mu)$  with a divergence $D_u(\nu \| \mu)$ (Theorem \ref{th:Ce} and \ref{th:Cr*}).

The paper is organized as follows. 
In section \ref{se:AB}, the main theorems about $k_n$ are stated. 
Section \ref{se:Cn} describes the results when no particular information theoretical structure is assumed.
The proofs are given in Section \ref{se:proof}.
Section \ref {se:con} concludes the paper.

\section{
Results for $ A_n $-code and $ B_n $-code
}
\label{se:AB}
A sequence
$
\X
=
\{ X^n\}_{n=1}^{\infty}   
$ 
of random variable $X^n$ is called a general source (cf. \cite{H03,HV}).
We consider an general source 
$(\X_1,\X_2)=\{ (X_1^n,X_2^n)\}_{n=1}^{\infty}$
where  $(X_1^n,X_2^n)$ is a random variable taking values in 
$\C X_1^n\times\C X_2^n$, here
the set $\C X_1^n$ is the source alphabet and the set $\C X_2^n$ is the side-information alphabet. 
We call $\X_1=\{ X_1^n\}$ and $\X_2=\{ X_2^n\}$ the information source and the side information, respectively. 
Let $ P(F)$ be the probability of event $F$.

\begin{figure}
 \includegraphics[width=6cm]{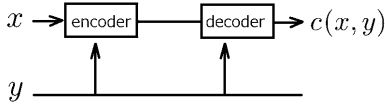}
	\caption{$c\in A_n$.}
	\label{fi:A-code}
\end{figure}

\begin{figure}
	\includegraphics[width=6cm]{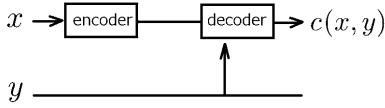}
	\caption{$c\in B_n$.}
	\label{fi:B-code}
\end{figure}

In this section, we study fixed-length source coding with side information. 
If the side information is accessible for both encoding and decoding, then encoding and decoding can be expressed as
$$
\varphi_n\,:\,\C X_1^n\times \C X_2^n \,\rightarrow\,\C M_n
$$
and
$$
\psi_n\,:\,\C M_n\times\C X_2^n\,\rightarrow\,\C X_1^n, 
$$
respectively, 
where $\C M_n =\{ 1,2,\cdots ,M_n\}$, here $M_n$ is a positive integer. 
Denote by $A_n$ the set of all pairs $c_n = (\varphi_n,\psi_n)$ of encoding and decoding. 
For simplicity, 
$c_n=(\varphi_n,\psi_n)\in A_n$
 is identified with the mapping 
$c_n\,:\,\C X_1^n\times \C X_2^n  \,\rightarrow\,\C X_1^n$ 
defined by 
$$
c_n(x,y)=\psi_n(\varphi_n(x,y),y), \quad
(x,y)\in \C X_1^n\times\C X_2^n ,
$$
(Fig. \ref{fi:A-code}).
Denote by 
\Eq
\| c_n\|=M_n
\label{eq:2.1}
\Qe 
the size of $c_n$. 
The rate of
$\{ c_n\}_{n=1}^{\infty}$ is defined by 
$$
\LSUPn \frac 1n\log \| c_n\| .
$$ 
We define the error probability of $c_n$ by
$$
\Ve_n = \Ve_n(c_n) = P ( c_n(X_1^n,X_2^n)\ne X_1^n ).
$$
Denote by
\Eq
T_n( c_n)=\{ (x,y)\in \C X_1^n\times\C X_2^n:\ 
c_n(x,y)=x
\} 
\label{eq:2.2}
\Qe
the set of correctly decoded $(x,y)$. 
Clearly, $\Ve_n=\Ve_n (c_n)=P(T_n(c_n)^c)$ holds.
If only decoding is accessible to the side information, 
then encoding and decoding can be expressed as
$$
f_n\,:\,\C X_1^n \,\rightarrow\,\C M_n
$$
and
$$
g_n\,:\,\C M_n\times\C X_2^n\,\rightarrow\,\C X_1^n,
$$
respectively. 
Denote by $B_n$ the set of all pairs $c_n = (f_n,g_n)$. 
For simplicity, 
we put $c_n(x,y)=g_n(f_n(x),y)$ for $c_n \in B_n$
(Fig. \ref{fi:B-code}).
Hereafter, we will call element $c=(\varphi ,\psi )$ of $A_n$ (resp. $B_n$) as $A_n$-code (resp. $B_n$-code).
Since for all  $c_n=(f_n,g_n)\in B_n$ there exists $(\varphi_n,\psi_n)\in A_n$ such that 
$$f_n(x)=\varphi_n(x,y),\quad \mbox { for all }(x,y)\in \C X_1^n\times \C X_2^n$$ and $g_n=\psi_n$, 
we have $B_n\subset A_n$.
That is, a $B_n$-code accessible to the side information only at the time of decoding can be regarded as an $A_n$-code accessible to the side information  in both encoding and decoding.

The minimum achievable coding rates of the $ A_n $-code and $ B_n $-code are defined as follows.

\Tlad

\begin{define}
A rate $R$ is said to be $A$-achievable (resp. $B$-achievable) if there exists 
$ c_n \in A_n$ (resp. 
\TS 
$B_n$) satisfying 
\Eq
\LLL \| c_n\| \le R
\label{eq:2.3}
\Qe
and
\Eq
\lim_{n\to\infty} \Ve_n(c_n) =0.
\label{eq:2.4}
\Qe
The minimum achievable coding rates $ R_A $ and $R_B$ are defined by
\Eq
R_A= \inf \{ R:\ \mbox{ $R$ is $A$-achievable}\}
\Qe
\Eq
R_B = \inf \{ R:\ \mbox{ $R$ is $B$-achievable}\} .
\Qe
\end{define}

\Tlad

\begin{define}
A rate $R$ is said to be $A^*$-achievable (resp. $B^*$-achievable) if there exists 
$ c_n \in A_n$ 
\TS
(resp. $B_n$) satisfying 
(\ref{eq:2.3}) and
\Eq
\LINFn \Ve_n (c_n) <1 .
\label{eq:2.7}
\Qe
The minimum achievable coding rates $ R_A^* $ and $R_B^*$ are defined by
\Eq
R_A^*= \inf \{ R:\ \mbox{$R$ is $A^*$-achievable}\}
\Qe
\Eq
R_B^*= \inf \{ R:\ \mbox{ $R$ is $B^*$-achievable}\} .
\Qe
\end{define}

\Tlad
 $R_B= \O{H}(\X_1|\X_2)$  can be obtained from Theorem 2 of  \cite{MK} by Miyake and Kanaya, where $\O{H}(\X_1|\X_2)$ is the conditional sup-entropy rate of $\X_1$ for given $\X_2$. 
Since $R_A=\O{H}(\X_1|\X_2)$ can be proved in the same way as the proof of this theorem, $R_A = R_B$ holds.
Similarly, it can be seen that $ R_A ^ * = R_B ^ * $ holds.
Summarizing the above, the following theorem is obtained.

\Tlad

\begin{theorem}
\label{th:MK}
\TS
\Eq
R_A =R_B, \quad R_A^*= R_B^* .
\label{eq:2.10}
\Qe
\end{theorem}

\Tlad
Theorem \ref {th:MK} means that the minimum acievable coding rate can not be reduced, even if the side information is accessible during encoding.
The purpose of this paper is to clarify the merits of being able to access the side information during encoding.
For this purpose, we investigate the number of $ B_n $-codes needed to get the same performance as an $ A_n $-code.

\begin{figure}
	\includegraphics[width=6cm]{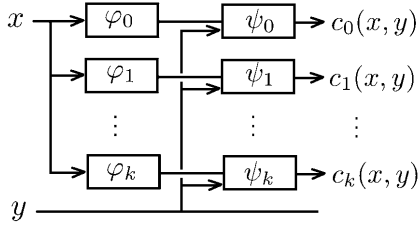}
	\caption{%
					$k+1$ $B_n$-codes. }
	\label{fi:kB-code}
\end{figure}

First, imagine a situation where $ k + 1 $ people encode and decode the same source separately.
Let $ c_ {0n}, c_ {1n}, \cdots, c_ {kn} $ be $ B_n $-codes that each person uses.
We will fix $ n $ for a while, so for simplicity we will write $ c_i = c_{in} $
(Fig. \ref{fi:kB-code}).
$T_n(c_i)$ is the event that the message is correctly decoded using $c_i$.
$
T_n(c_0)\cup T_n(c_1)\cup \cdots \cup T_n(c_k)
$
 is the event that a message has been correctly decoded with at least one $ c_i $.
Let 
$$
\Ve_n (c_0, c_1, \cdots ,c_k)
=
 P(T_n(c_0)^c\cap T_n(c_1)^c\cap \cdots \cap T_n(c_k)^c)
$$
be the probability that a message is decoded incorrectly by 
all $c_i$ ($i=0,1,\cdots ,k$).
Next, we will explain how to use the above quantities to determine the number of $ B_n $-codes that have the same performance as an $ A_n $-code.
For each $c\in A_n$, let $ K_n (c) $ be the smallest non-negative integer k such that there exists $ c_i \in B_n $ ($ i = 0, 1, \cdots, k $)
satisfying 
$\| c\| =\| c_i\|$
and 
\Eq
\Ve_n(c)
\ge
\Ve_n(c_0, c_1, \cdots ,c_k). 
\label{eq:2.11}
\Qe
In the case of $K_n(c)=0$, there exists $c_0\in B_n$
such that 
$\| c\| =\| c_0\| $
and
$
\Ve_n(c)
\ge
\Ve_n (c_0).
$
Since $ c $ and $ c_0 $ are the same size, and the error probability of $ c $ is not less than that of $ c_0 $, it can not be said that $ c $ is better than $ c_0 $.
Even though $ c $ can access the side information during encoding, $ c $ performance can not be better than inaccessible code $ c_0 $.
Thus, if $ K_n (c) = 0 $, the $A_n$-code $ c $ can not take full advantage of the ability to access the side information during encoding.
In the case of $K(c)\ge 1$, for any $c_0\in B_n$ satisfying $\| c\| =\| c_0\| $, we have
$
\Ve_n(c)
<
\Ve_n (c_0). 
$
Therefore, it can be said that the error probability of the code $ c $ could be reduced by accessing the side information.
Let us consider the case of $ K_n (c) = k $.
In this case, 
\Eq
\Ve_n(c)
<
\Ve_n(c_0, c_1, \cdots ,c_{k-1})
\label{eq:2.12}
\Qe
holds for any $c_0,c_1, \cdots , c_{k-1}\in B_n$ satisfying $\| c\| =\| c_i\| $. 
The sizes of $ c $ and $ c_i $ are the same, and the error probability of $ c $ is smaller than the probability that a message is decoded incorrectly by all $c_i$ ($i=0,1,\cdots ,k-1$).
In this sense, it can be said that $ c $ performs better than 
$k$ $B_n$-codes.
However, since there are $k + 1$ $ B_n$-codes satisfying
(\ref{eq:2.11}), it can not be said that the performance is better than 
$k + 1$ $ B_n $-codes.
Therefore, the performance of $c$ is considered to be greater than $k$ $B_n$-codes but less than or equal to $k+1$ $B_n$-codes. 
Since the number of $B_n$-codes is a natural number, we consider that the performance of c is equivalent to that of $k+1$ $B_n$-codes, where $k=K_n(c)$. 
It does not seem necessary to determine the value of $ K_n (c) $ for 
arbitlary $ c \in A_n$.
The important thing is to determine the value of $ K_n (c ^ *) $ when there is the optimal $A_n$-code $ c ^ * $ (the code with the smallest error probability among codes of the same size).
For this purpose the following quantities are introduced:
the minimum error probability $e_A (M_n)$ is defined by
\Eq
e_A(M_n)=\min_{c\in A_n}\{ \Ve_n(c): \| c\|=M_n \} .
\label{eq:2.13}
\Qe
$e_A (M_n) $ is the minimum value of the error probability of $ A_n $-code of size $ M_n $.
Similarly, for non-negative integers $k_n$,
the minimum error probability $e_B(M_n; k_n)$ is defined by
$$
e_B(M_n; k_n)=\min_{c_i\in B_n}\{ \Ve_n(c_0, \cdots ,c_{k_n}): \| c_i\|=M_n, 
0\le i \le k_n \}
$$
$ e_B (M; k_n) $ is the minimum value of the probability that all $ k_n + 1 $ $ B_n $-codes of size $ M_n $ are erroneously decoded.
If $ k_n = 0 $ then $ e_B (M_n; 0) $ is the minimum value of the error probability of $ B_n $-code of size $ M_n $.
$ e_B (M; k_n) $ is a quantity representing the performance of $ k_n + 1 $ $ B_n $-codes. By definition, $ k_n = K_n (c) $ and $ e_B (M; k_n) \le \Ve_n (c) <e_B (M; k_n-1) $ are equivalent.
Since the optimal $A_n$-code $c^ *$ of size $M_n$ satisfies $ e_A (M_n) = \Ve_n (c^*) $, it turns out that 
$$ k_n = K_n (c ^ *) 
$$ 
and
\Eq
e_B(M_n; k_n) \le e_A(M_n) < e_B(M_n; k_n-1)
\label{eq:2.14}
\Qe 
are equivalent.
Therefore, we conclude that the performance of the optimal $ A_n $-code is equivalent to that of $ k_n + 1 $ $ B_n $-codes when (\ref {eq:2.14}) holds.
The magnitude of the advantage of being able to access the side information during encoding is characterized by $k_n$ satisfying (\ref {eq:2.14}).

Now we will consider the same problem in asymptotic situations where $n\to\infty$.
To state the problems precisely, we introduce two conditions
\Eq
\LLL\|c_n\| \le R ,\quad c_n\in A_n
\label{eq:2.15}
\Qe
and 
\Eq
\LLL \left (\max_{0\le i\le k_n} \|c_{in}\| \right ) \le R, \quad 
 c_{in}\in B_n
\label{eq:2.16}
\Qe
on the coding rates. 
Thereafter, when considering the condition (\ref {eq:2.15}), $ \Ve_n = \Ve_n (c_n) $, and when considering the condition (\ref {eq:2.16}) $ \Ve_n = \Ve_n (c_ {0n}, \cdots, c_ {k_nn}) $.
Let $\Ve \ge 0$. 
Since the limit of $\lim_{n\to\infty} \Ve_n $ as $n\to\infty$ may not exist in general, we need to consider a condition 
\Eq
\LSUPn \Ve_n 
\le \Ve
\label{eq:2.17}
\Qe
concerning the error probability. 

\Tlad

\begin{define} 
A rate $R$ is said to be $\Ve$-achievable if there exists 
$ c_n \in A_n$ satisfying 
(\ref{eq:2.15}) and (\ref{eq:2.17}). 
\TS
A rate $R$ is said to be $(\Ve ;{k_n} )$-achievable if there exists 
$c_{in}\in B_n$ satisfying 
(\ref{eq:2.16}) and (\ref{eq:2.17}). 
The minimum error probabilities $\Ve_A(R)$ and $\Ve_B(R; {k_n} )$ are defined by
\Eqas
& &
\Ve_A(R)=\inf\{  \Ve: 
\mbox{ $R$ is $\Ve$-achievable } \} ,
\\
& &
\Ve_B(R; {k_n} )
=\inf \{  \Ve: 
\mbox{ $R$ is $(\Ve ;{k_n} )$-achievable } 
\} .
\Qeas
\end{define}

$ \Ve_A (R) $ (resp. $\Ve_B (R; {k_n}) $) is the upper limit of error probability $\Ve_n$ of the optimal $ A_n $-code (resp. $ k_n+1 $ $ B_n $-codes) under the constraint that the coding rate is less than $ R $. 
Note that $\Ve_B (R; {k_n}) $  is a quantity that does not depend on $n$.
The important thing is to find $ k_n $ which satisfies the condition $ \Ve_A (R) = \Ve_B (R; {k_n}) $. 
We have the following theorem on $k_n$. 

\Tlad

\begin{theorem}
\label{th:ABe}
Let $s, R>0$ be arbitrary. Then, for any non-negative integer $k_n\le n^s$ ($n=1,2,\cdots$), 
\TS
\Eq
\Ve_A(R)=\Ve_B(R; k_n) .
\label{eq:2.18}
\Qe
\end{theorem}

The proof is given in section \ref {se:proof}.
Note that (\ref {eq:2.18}) holds for $ k_n = 0 $ ($ n = 1, 2, \cdots $).
This means that the performance of the optimal $ A_n $-code corresponds to that of one $ B_n $-code.
Although the $ A_n $-code can access the side information during encoding, it can not be said that the optimal $ A_n $-code is better than one $ B_n $-code.
Therefore, it is not possible to reduce the upper limit of error probability by making the side information accessible during encoding.

\Tlad

\begin{remark}
\label{re:ABe}
We would like to explain the practical meaning of Theorem \ref{th:ABe}.
First, we explain that 
\TS
$k_n+1$ $B_n$-codes are useful for noiseless transmission.
Suppose a sender wants to send message $X_1^n$ over a noiseless channel, where a receiver can access $X_2^n$ correlated with $X_1^n$, but a transmitter can not access $X_2^n$.
When  $\varphi_n(X_1^n)$ is input to the noiseless channel, the output is also  $\varphi_n(X_1^n)$, and the receiver decodes it to obtain $\psi_n(\varphi_n(X_1^n), X_2^n)$.
Therefore $c=(\varphi_n, \psi_n)\in B_n$.
Similarly, let us consider one sender wants to send message $X_1^n$ to $ k_n + 1 $ recipients using noiseless channels.
\begin{figure}
	\centering
	\includegraphics[width=6cm]{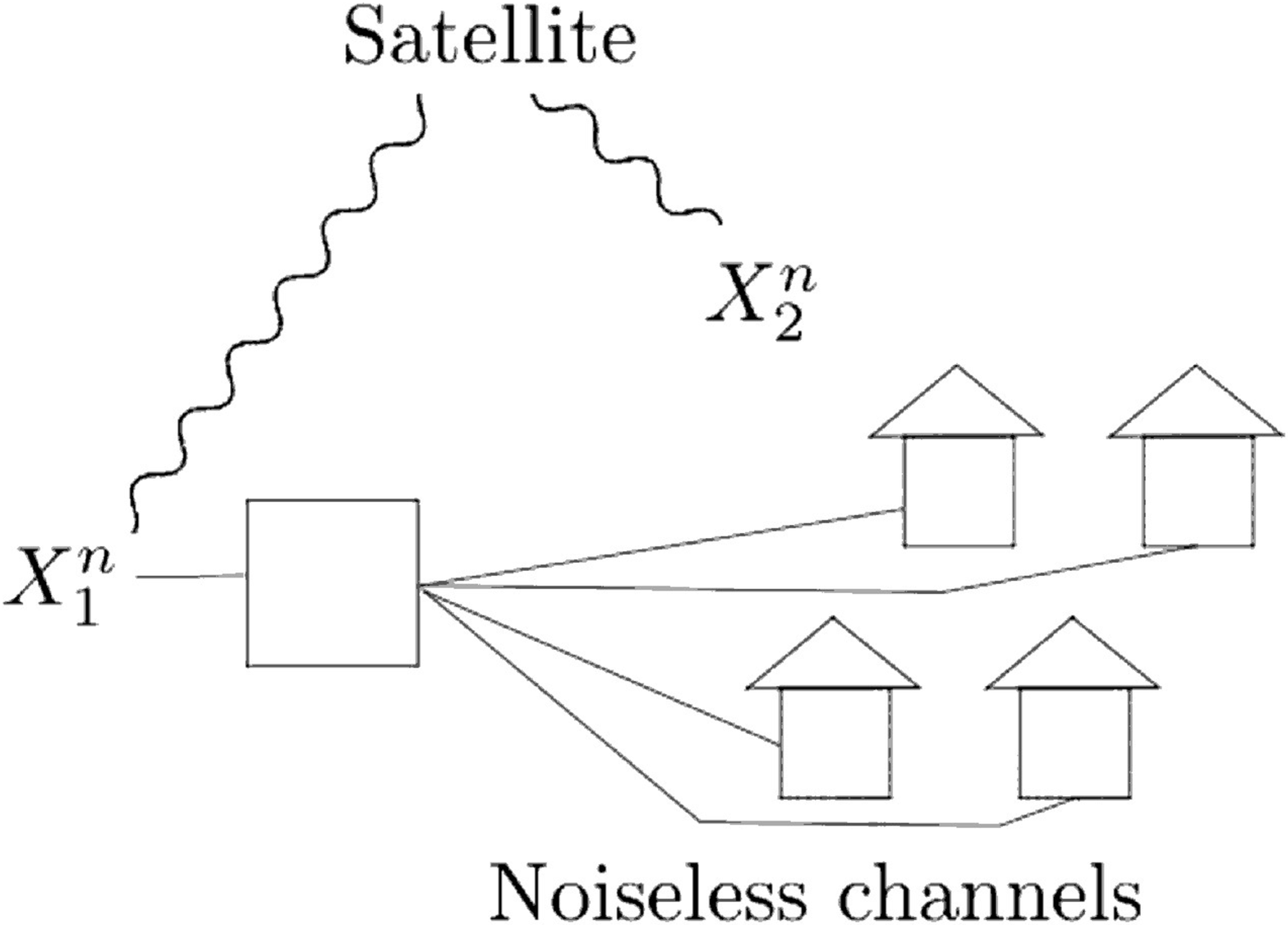}
	\caption{Example of sending $X_2^n$ by satellite in advance. ($k_n=3$)}
	\label{fi:satellite}
\end{figure}
Suppose each recipient uses a noiseless channel privately and does not cooperate with each other, but all recipients have access to $ X_2 ^ n $ (Fig. \ref{fi:satellite}).
When we want to reduce the probability that all decoders will cause errors, the transmitter should send different data $\varphi_{in}(X_1^n)$ to each receiver $i$ ($i=0,1,\cdots ,k_n$). 
Therefore, let us use different $B_n$-codes $c_i=(\varphi_{in}, \psi_{in})$ for each noiseless channel.
Then, the probability that all decoders will cause errors is $\Ve_n(c_0, c_1, \cdots ,c_{k_n})$. 
Next, we will explain Theorem  \ref{th:ABe}.
Theorem  \ref{th:ABe}  means $\Ve_B(R; k_n)$ is constant for all $k_n\le n^s$. 
Therefore $\Ve_B(R; [n^s]) =\Ve_B(R; 0)$. 
That is, roughly speaking, 
\[
\Ve_n(c_0^*, c_1^*, \cdots ,c_{[n^s]}^*) 
\simeq
\Ve_n(c_0^*)
\]
holds for the optimal $B_n$-codes $c_i^*=(\varphi_{in}^*, \psi_{in}^*)$  in asymptotic situations where $n$ is increased.
On the other hand, since from the definition of $\Ve_n(c_0, \cdots ,c_k)$ we have $\Ve_n(c, c \cdots ,c) =\Ve_n (c)$, it follows that
\[
\Ve_n(c_0^*, c_1^*, \cdots ,c_{[n^s]}^*) 
\simeq
\Ve_n(c_0^*,c_0^* \cdots ,c_0^*) .
\]
This means that in the asymptotic situation, sending different data  $\varphi_{in}^*(X_1^n)$  to each receiver $i$ can not reduce the probability that all decoders will cause errors. 
Thus, for at most  $ n ^ s $  receivers, it is recommended that the transmitter send the same data $\varphi_{0n}^*(X_1^n)$ to all receivers and that  $\psi_{0n}^*$ be used as the decoder for each receiver $i$.
\Prend
\end{remark}

\Tlad

Next, we consider the convergence rate of the error probability.
Depending on the value of $ R $, we consider the high-rate case and the low-rate case separately.

\subsection{
High-rate case}

When the coding rate $ R $ is sufficiently large, the error probability of the optimal $ A_n $-code converges zero exponentially fast.
Therefore, we will evaluate the encoder performance at the exponential rate (so called error exponent). 
To state the problems precisely, we introduce a condition
\Eq
\lll \frac 1
{\Ve_n}\ge r
\label{eq:2.19}
\Qe
concerning the error exponent.

\Tlad

\begin{define}
A rate $R$ is said to be $r$-achievable if there exists 
$ c_n \in A_n$ satisfying 
(\ref{eq:2.15}) and (\ref{eq:2.19}). 
\TS
A rate $R$ is said to be $(r ;{k_n} )$-achievable if there exists 
$c_{in}\in B_n$ satisfying 
(\ref{eq:2.16}) and (\ref{eq:2.19}). 
The reliability functions $\rho_A (R)$ and $\rho_B (R; {k_n} )$ are defined by  
\Eqas
& &
\rho_A (R)=\sup\{  r: 
\mbox{ $R$ is $r$-achievable } \} , 
\\
& &
\rho_B (R; {k_n} )
=\sup \{  r: 
\mbox{ $R$ is $(r ;{k_n} )$-achievable } 
\} .
\Qeas
\end{define}

The following theorem is the main result of this paper.

\Tlad

\begin{theorem}
\label{th:ABr}
Let $R>0$ be a continuous point of $\rho_A (\cdot )$. Then, 
for any $s>1$
\TS
\Eq
\rho_A (R)=\rho_B (R; k_n ), 
\label{eq:2.20}
\Qe
where $k_n=[n^s]$. 
\end{theorem}

The proof is given in section \ref {se:proof}.
It can be seen from (\ref {eq:2.20}) that the error exponent of the optimal 
$ A_n $-code is equal to that of $ [n ^ s] + 1 $ $B_n $-codes.
Therefore, in the high-rate case, we consider that the performance of the optimal $ A_n $-code  is equivalent to that of $ [n ^ s] + 1 $ $ B_n $-codes.

\subsection{
Low-rate case
}
If the coding rate $ R $ is sufficiently small, then the error probability tends to one.
Therefore, we will evaluate the encoder's performance at the speed at which the probability of being correctly decoded converges to zero.
To state the problems precisely, we introduce a condition
\Eq
\LLL \frac 1
{1-\Ve_n}
\le 
r 
\label{eq:2.21}
\Qe
on the probability of being correctly decoded. 

\Tlad

\begin{define}
A rate $R$ is said to be $r^*$-achievable if there exists 
$ c_n \in A_n$ satisfying 
(\ref{eq:2.15}) and (\ref{eq:2.21}). 
\TS
A rate $R$ is said to be $(r^* ;{k_n} )$-achievable if there exists 
$c_{in}\in B_n$ satisfying 
(\ref{eq:2.16}) and (\ref{eq:2.21}). 
The reliability functions $\rho_A^* (R)$ and $\rho_B^* (R; {k_n} )$ are defined by  
\Eq
\rho_A^* (R)=\inf\{  r: 
\mbox{ $R$ is $r^*$-achievable } \} ,
\Qe
\Eq
\rho_B^* (R; {k_n} )
=\inf \{  r: 
\mbox{ $R$ is $(r^* ;{k_n} )$-achievable } 
\} .
\Qe
\end{define}

We have the following theorem. 

\Tlad

\begin{theorem}
\label{th:ABr*}
Let $s , R>0$ be arbitrary. Then, for any non-negative integer $k_n\le n^s$ ($n=1,2,\cdots$), 
\TS
\Eq
\rho_A^* (R)=\rho_B^* (R; {k_n} ). 
\label{eq:2.24}
\Qe
\end{theorem}
Note that (\ref{eq:2.24}) holds for $ k_n = 0 $ ($ n = 1, 2, \cdots $).
Therefore, in the low-rate case, we consider that the performance of the optimal $ A_n $-code corresponds to that of one $ B_n $-code.

\section{
Results for $(C_n ,\| \cdot \| , T_n)$ which is generalized $A_n$-code and $B_n$-code.
}

\label{se:Cn}

In this section, in order to prove theorems in the previous section, we define $(C_n ,\| \cdot \| , T_n)$ as a general theory of probability theory without assuming any particular information theoretical structure.
The results of this section can be applied to the $ A_n $-code (resp. $ B_n $-code) by placing $ C_n = A_n $ (resp. $ C_n = B_n $).

Let $ \C X_n $ be a set, and $ 2 ^ {\C X_n} $ be a power set of $ \C X_n $.
Let us consider a triple  $(C_n,\| \cdot \| ,T_n)$, where $C_n$ is a set, $\| \cdot \| $ is a non-negative function on $C_n$, and $T_n$ is a set-valued function $T_n: C_n\to 2^{\C X_n}$. 
Note that  $ T_n (c) \subseteq \C X_n $ for all $c\in C_n$. 
Since we are interested in the probability of $ T_n (c) $, we will consider a measurable space $ (\Omega_n, \C F_n) $ satisfying $\{ T_n(c):\ c\in C_n\} \subset \C F_n$. 
Denote by $M(\Omega_n)$ the set of all probability measures on 
$ (\Omega_n, \C F_n) $, and denote by $\C M(\Omega)$ the set of all sequences $\mu=\{ \mu_n\}_{n=1}^{\infty}$ satisfying $\mu_n\in M(\Omega_n)$ ($n=1,2,\cdots $).

\Tlad

\begin{define}\label{de:1} 
We say that $(C_n, \| \cdot \| , T_n)$ is (finitely) subadditive if for all $c_1,c_2\in C_n$ there exists 
\TS
$c\in C_n$ such that 
$$\| c \|\le \| c_1 \|+\| c_2 \|$$
and 
$$T_n(c)=T_n(c_1)\cup T_n(c_2).$$
\end{define}

\Tlad

\begin{remark} 
When we want to consider  $ A_n $-code (resp. $ B_n $-code), 
we put $ C_n = A_n $ (resp. $ C_n = B_n $) 
\TS
and $\C X_n = \C X_1 ^ n \times \C X_2 ^ n $, define $ \| \cdot \| $ by (\ref {eq:2.1}), and define $T_n(\cdot )$ by (\ref{eq:2.2}). 
The probability measure of $(X_n^1, X_n^2)$ is $\mu_n\in M(\Omega_n)$.
The sample space $\Omega_n$ may be anything that satisfies $\C X_n \subseteq \Omega_n$.
Of course, it may be $\Omega_n =\C X_n$, but it may be  $\Omega_n \ne \C X_n$.
The difference between $ A_n $-code and $ B_n $-code is whether it is subadditive.
Although details will be described later, $A_n$-code is subadditive iProposition \ref{pr:an}j. 
On the other hand, it is not known whether $B_n$-code is subadditive. 
\Prend 
\end{remark}

\Tlad

In this section, without assuming any information-theoretic structure, we introduce quantities corresponding to  minimum achievable coding rates and reliability functions. 
For this purpose, we denote by $\C{C}$ the set of all sequences $\B c=\{ c_n\}_{n=1}^{\infty}$ ($c_n\in C_n$). 
It is useful to introduce the following subset of $\C{C}$: 
\Eqa
\C C(R)
&=&
\left\{
\B c =\{ c_n\}\in \C C:\ 
\LLL \| c_n \| 
\le 
R
\right \}
\label{eq:3.25}
\\
\C C_T(\mu )
&=&
\left\{
\B c =\{ c_n\}\in \C C:\ 
\lim_{n\to\infty} \mu_n(T_n(c_n)) =1
\right\} 
\label{eq:3.26}
\\
\C C_T^*(\mu )
&=&
\left\{
\B c =\{ c_n\}\in \C C:\ 
\LINFn \mu_n(T_n(c_n)^c) <1
\right\} .
\Qea
The quantities $R_C(\mu )$ and $R_C^*(\mu )$ corresponding to the minimum achievable coding rate are defined as follows.

\begin{define} 
\TS
\Eqa
R_C(\mu )
&=&
\inf_{\B c \in \C C_T(\mu )}
\LLL\| c_n \| 
\label{eq:3.28}
\\
R_C^*(\mu )
&=&
\inf_{\B c \in \C C_T^*(\mu )}
\LLL\| c_n \| . 
\label{eq:3.29}
\Qea
\end{define}

We would like to define similar quantities as defined for $k_n+1$ $B_n$-codes in the previous section. 
Denote by $\C C(\cdot ;k_n)$ the set of all sequence $\B c =\{ (c_{0n}, c_{1n}, \cdots , c_{k_nn})\}_{n=1}^{\infty}$ ($c_{in}\in C_n$, $i=0,1,\cdots , k_n$). 
It is useful to introduce the following subset of $\C C(\cdot ;k_n)$: 
\Eqa
\C C(R;k_n)
&=&
\left \{
\B c =\{ (c_{0n}, \cdots , c_{k_nn}) \} \in \C C(\cdot ;k_n) :\ 
\limsup_{n\to\infty}\frac 1n\log \left (\max_{ 0\le i\le k_n}\| c_{in} \| \right )
\le
R
\right\} . 
\label{eq:3.30}
\Qea
Especially if $k_n = 0$ ($n = 1, 2, \cdots $) then $ \C C (\cdot; k_n) = \C C $ and  $\C C(R;k_n)=\C C(R)$. 
The quantities corresponding to $\Ve_A(R)$ and $\Ve_B(R; k_n)$ are defined as follows.

\begin{define} 
\TS
\Eq
\Ve_C(R| \mu )
=
\inf_{\B c \in \C C(R)}
\LSUPn \mu_n(T_n(c_n)^c)  
\label{eq:3.31}
\Qe
\Eq
\Ve_C(R| \mu ;k_n)
=
\inf_{\B c \in \C C(R;k_n)}
\LSUPn \mu_n\left (\bigcap_{i=0}^{k_n} T_{n}(c_{in})^c \right ).   
\label{eq:3.32}
\Qe
\end{define}

Clearly $0\le \Ve_C(R| \mu ;k_n) \le \Ve_C(R|\mu )\le 1$. 
Especially if $k_n = 0$ ($n = 1, 2, \cdots $) then $\Ve_C(R| \mu ; k_n) =\Ve_C(R| \mu )$.

Similarly, the quantities  corresponding to the reliability functions are defined as follows.

\Tlad

\begin{define} 
\TS
\Eqa
\rho_C(R|\mu ) 
&=&
\sup_{\B c\in \C C(R)}
\liminf_{n\to\infty}\frac 1n\log 
\frac 1{\mu_n(T_n(c_n)^c)}
\label{eq:3.33}
\\
\rho_C(R|\mu ;{k_n} )
&=&
\sup_{\B c \in C(R;k_n)}
\liminf_{n\to\infty}\frac 1n\log \frac 1{
\mu_n\left (\bigcap_{i=0}^{k_n} T_{n}(c_{in})^c \right )}
\label{eq:3.34}
\Qea
\Eqa
\rho_C^*(R|\mu ) 
&=&
\inf_{\B c\in \C C(R)}
\limsup_{n\to\infty}\frac 1n\log 
\frac 1{\mu_n(T_n(c_n))}
\label{eq:3.35}
\\
\rho_C^*(R|\mu ;{k_n} )
&=&
\inf_{\B c \in C(R;k_n)}
\limsup_{n\to\infty}\frac 1n\log \frac 1{
\mu_n\left (\bigcup_{i=0}^{k_n} T_{n}(c_{in}) \right )}
\label{eq:3.36}
\Qea
\end{define}

Clearly 
$0\le \rho_C(R|\mu ) \le \rho_C(R|\mu ;{k_n} )$ and 
$ \rho_C^*(R|\mu ) \ge \rho_C^*(R|\mu ;{k_n} ) \ge 0$. 
Especially if $k_n = 0$ ($n = 1, 2, \cdots $) then $\rho_C (R|\mu ; k_n )=\rho_C (R|\mu)$ and
$\rho_C^* (R|\mu ; k_n )=\rho_C^* (R|\mu)$.

\Tlad

We define the following quantities needed to represent the results in this section.
For probability measures $\mu,\nu\in M(\Omega_n)$, 
the divergence of $\nu$ with respect to $\mu$ is defined as follows:
if $\nu\ll \mu$ then
$$
D(\nu\| \mu )
=
\int_{\Omega_n}  \log \frac{d\nu}{d\mu} (x) d\nu (x),
$$
otherwise set $D(\nu\|\mu)=\infty$.
For $\mu=\{ \nu_n\}, \nu=\{ \nu_n\} \in \C M(\Omega)$, 
we define the upper divergence
$D_u(\nu\|\mu)$ and the lower divergence $D_l(\nu\|\mu )$ by
\Eqas
D_u(\nu\|\mu)
&=&
\LSUPn \frac 1n D(\nu_n\| \mu_n)
\\
D_l(\nu\|\mu )
&=&
\LINFn \frac 1n D(\nu_n\| \mu_n) .
\Qeas
Similarly we define $G(\nu \|\mu )$, $G_u(\nu\|\mu )$ and $G_l(\nu\|\mu )$ as follows. 
if $\nu\ll \mu$ then 
\Eq
G(\nu \|\mu )
=
\int_{\Omega_n}\frac{d\nu_n}{d\mu_n}(x)d\nu_n(x),
\Qe
otherwise set $G(\nu \|\mu )=\infty$. 
We put
\Eq
G_u(\nu\|\mu )
=
\LSUPn G(\nu_n \|\mu_n )
\Qe
\Eq
G_l(\nu\|\mu )
=
\LINFn G(\nu_n \|\mu_n ). 
\Qe

Now, we state our main results in this section. 

\Tlad

\begin{theorem} 
\label{th:Ce}
If $\Ve_C(R|\mu ) \ne 1$ then
\TS
\Eq
\frac 1{1-\Ve_C(R|\mu )}
=
\inf_{\nu \in \C M(\Omega ) } \{ G_u(\nu\|\mu ):\ R_C(\nu )\le R\} . 
\label{eq:3.40}
\Qe
\end{theorem}

If $\Ve_C(R|\mu )=1$, a rate $R$ is a small value, so it is important to determine the value of $\rho_C^*(R|\mu )$.
The following result for $ \rho_C ^ * (R | \mu) $ is substantially the same as Theorem 4 of \cite{Ir00}.

\Tlad

\begin{theorem}[cf. \cite{Ir00}] 
\label{th:Cr*}
For any $\mu\in \C M(\Omega)$ and any real number $R$, 
\TS
\Eq
\rho_C^*(R|\mu ) 
=
\inf_{\nu\in \C M(\Omega )}\{ D_u(\nu\| \mu):\, R_C(\nu)\le R\} .
\label{eq:3.41}
\Qe
\end{theorem}

\Tlad

Interestingly, the difference between the right-hand side of (\ref {eq:3.40}) and that of (\ref {eq:3.41}) is only $G_u(\nu\|\mu )$ and $D_u(\nu\|\mu)$.
Therefore, we consider that $ G_u (\nu \|\mu) $ is an important quantity as well as divergence.
The following lemma holds for quantities $\Ve_C (R|\mu;{k_n})$, $\rho_C (R|\mu ;{k_n} )$ and $\rho_C^*(R|\mu ; k_n )$.

\Tlad

\begin{lemma} 
\label{le:r*kn}
For any $\mu\in \C M(\Omega)$, any real number $R$, $s$, and any non-negative integer $k_n\le n^s$ 
\TS
($n=1,2,\cdots$), 
\Eq
\rho_C^* (R|\mu) 
=
\rho_C^* (R|\mu ;{k_n} ) . 
\label{eq:3.42}
\Qe
If $(C_n, \| \cdot \| , T_n)$ is subadditive then 
\Eq
\Ve_C (R|\mu) = \Ve_C (R|\mu ;{k_n} ) 
\label{eq:3.43}
\Qe
\Eq
\rho_C (R|\mu) = \rho_C (R|\mu ;{k_n} ) .
\label{eq:3.44}
\Qe
\end{lemma}

\Tlad

The following result is substantially the same as the result for the first inequality of (11) of \cite{IrIh}.

\begin{lemma}[cf. \cite{IrIh}] 
\label{le:main0}
For any $\mu\in \C M(\Omega)$ and any real number $R$, 
\TS
\Eq
\rho_C (R|\mu) 
\le
\inf_{\nu\in \C M(\Omega )}\{ D_l(\nu\| \mu):\, R_C^*(\nu)> R\} .
\label{eq:3.45}
\Qe
\end{lemma}

In order to express the results for $\rho_C (R|\mu)$, we introduce the following conditions.
\Eq
\label{eq:3.46}
\mbox{ There exists $\B \nu \in \C M(\Omega )$ such that $R_C^*(\B \nu )\ge R$.}
\Qe

\Tlad

\begin{remark} 
Let us explain that the condition (\ref{eq:3.46}) is satisfied when the sample space  $\Omega_n$ is sufficiently 
\TS
large. 
For a sufficiently large  $\Omega_n$, there exists $e\in \Omega_n$ such that $e\not\in \C X_n$.
This means that there exists a sequence $\B \nu =\{ \nu_n\} \in \C M(\Omega )$ satisfying  $\nu_n(\C X_n^{\ c})=1$.
Since $\nu_n(T_n (c_n)^c)\ge \nu_n(\C X_n^{\, c}) =1$ for all $c_n\in C_n$, we have  $\C C_T^*(\nu )=\emptyset$. 
Therefore $ R_C ^ * (\nu) = \infty \ge R $. 
\Prend
\end{remark}

\Tlad

The following lemma plays an important role in deriving the main results of this paper.

\Tlad

\begin{lemma}
\label{le:k2}
\rm
Suppose that the condition (\ref {eq:3.46}) is satisfied. 
Let $s>1$ be arbitrary. Then, for any 
\TS
$\mu\in \C M(\Omega)$ and any real number $R$, 
\Eq 
\rho_C (R|\mu ;{k_n} ) 
\ge
\inf_{\nu\in \C M(\Omega )}\{ D_l(\nu\| \mu):\, R_C^*(\nu)\ge R\}  ,
\Qe
where $k_n=[n^s]$ ($n=1,2,\cdots$). 
\end{lemma}

\Tlad

The following theorem is one of the main results of this paper.

\Tlad

\begin{theorem}
\label{th:Cr}
\rm
Suppose that the condition (\ref {eq:3.46}) is satisfied, and 
$(C_n, \| \cdot \| , T_n)$ is subadditive. 
Then, 
\TS
for any $\mu\in \C M(\Omega)$ and any real number $R$ 
\Eq 
\inf_{\nu\in \C M(\Omega )}\{ D_l(\nu\| \mu):\, R_C^*(\nu)\ge R\} 
\leq 
\rho_C (R|\mu) 
\le
\inf_{\nu\in \C M(\Omega )}\{ D_l(\nu\| \mu):\, R_C^*(\nu)> R\}  ,
\label{eq:3.48}
\Qe
with equalities if $\rho_C (\cdot |\mu)$ is continuous at $R$.
\end{theorem}

{\it Proof: }
The proof is obtained immediately from Lemma \ref{le:r*kn}, \ref{le:main0} and \ref{le:k2}. 
\Prend

\Tlad 
Theorem \ref{th:Cr} is a generalization of Theorem 3 of \cite{IrIh} concerning reliability function in the high-rate case.

\section{Proofs}

\label{se:proof}

\subsection{
Proofs of results for $(C_n,\| \cdot \|, T_n)$
}

The following two lemmas play a key role in proving the theorems.
For each $B_n\in \C F_n$ denote by $M(B_n)$ the set of all $\nu\in M(\Omega_n)$ satisfying $\nu (B_n)=1$, 
and for each $\B{B} = \{ B_n \}_{n=1}^{\infty}$ 
($B_n \in \C{F}_n$) 
denote by $\C{M}(\B{B})$ the set of all sequences 
$\nu=\{ \nu_n\} \in \C{M}(\C{X})$ 
satisfying $\lim_{n\rightarrow\infty}\nu_n(B_n)=1$. 
Proposition 1 of \cite {Ir00} give the following lemma. 

\Tlad

\begin{lemma}
\label{le:1}
\rm
For any $\mu = \{ \mu_n \} \in \C M(\Omega )$ 
and 
$\B B = \{ B_n \}$ ($B_n\in \C F_n$, $B_n\ne \emptyset$), 
\TS
\Eqa
\LLL \mu_n(B_n)
&=& -\min_{ \nu \in \C M(\B B)} D_l(\nu\|\mu ) 
\label{eq:4.49}
\\
\lll \mu_n(B_n)
&=&-\min_{ \nu \in \C M(\B B)} D_u(\nu\|\mu ) 
\label{eq:4.50} .
\Qea
\end{lemma}

\Tlad

\begin{remark}
Kuzuoka \cite{Ku18}  pointed out that the condition $ B_n \ne \emptyset $ is necessary for the equation (\ref {eq:4.50}) 
\TS
to hold.
Since this condition is missing in Proposition 1 of \cite{Ir00}, Lemma \ref{le:1} is given as a modified version. 
Here we emphasize that this lemma is also useful even if $ B_n = \emptyset$.
Without loss of generality, if $ \Omega_n $ is a sufficiently large set, there exists $ e \in \Omega_n $ such that $ \mu_n (e) = 0 $.
We put 
$$
B_n' 
=
\begin{cases}
e,\quad \mbox{ if $B_n = \emptyset$} \\
B_n,\quad \mbox{otherwise} \\
\end{cases}
$$ 
Then, since  $\mu_n(B_n') =\mu_n(B_n)$, we may identify set $B_n'$  with $B_n$.
By applying Lemma \ref {le:1} to $\B B'=\{ B_n'\}$, the value of 
\[
\lll \mu_n(B_n)
\]
can be determined even if $B_n=\emptyset$. 
\Prend
\end{remark}

\Tlad

Similar to the above lemma, the following Lemma holds.

\Tlad

\begin{lemma}
\label{le:lem2}
For any $\mu = \{ \mu_n \} \in \C M(\Omega )$ 
and 
$\B B = \{ B_n \}$ ($B_n\in \C F_n$, $B_n\ne \emptyset$), 
\TS
\Eq
\LINFn \frac 1{\mu_n(B_n)} 
=
\min_{\nu\in \C M(\B B)}G_l(\nu\|\mu )
\label{eq:4.51}
\Qe
\Eq
\LSUPn \frac 1{\mu_n(B_n)} 
=
\min_{\nu\in \C M(\B B)}G_u(\nu\|\mu ). 
\label{eq:4.52}
\Qe
\end{lemma}

The proof is given in the appendix.
\Tlad

{\it Proof of Theorem \ref{th:Ce}: }
From 
(\ref{eq:3.31})
we have
\Eqa
\frac 1{1-\Ve_C(R|\mu )}
&=&
\inf_{ \{ c_n\} \in \C C(R)}
\LSUPn
\frac 1{1-\mu_n (T_n(c_n)^c)}
\No
&=&
\inf_{ \{ c_n\} \in \C C(R)}
\LSUPn
\frac 1{\mu_n (T_n(c_n))}
\No
&=&
\inf_{ \{ c_n\} \in \C C(R)}
\left\{\LSUPn
\frac 1{\mu_n (T_n(c_n))}:\ T_n(c_n)\ne\emptyset \mbox{ for all }n\right \} .
\label{eq:4.53}
\Qea
Applying Lemma \ref{le:lem2} for $B_n=T_n(c_n)\ne \emptyset$, 
we have
\Eq
\LSUPn
\frac 1{\mu_n (T_n(c_n))}
=
\min_{\nu =\{ \nu_n\}} 
\left\{ 
G_u(\nu\|\mu ):\ \lim_{n\to\infty} \nu_n(T_n(c_n))=1
\right\} .
\label{eq:4.54}
\Qe
It follows from
(\ref{eq:4.53}), (\ref{eq:4.54}), (\ref{eq:3.25}), (\ref{eq:3.26}), and (\ref{eq:3.28}) 
that 
\Eqas
\frac 1{1-\Ve_C(R|\mu )}
 &=&
 \inf_{ \{ c_n\}\in \C C(R)}\min_{\nu}
\left \{ 
G_u(\nu\|\mu ):\  \lim_{n\to\infty} \nu_n(T_n(c_n))=1
\right \}
 \No
 &=&
\inf_{ \{ c_n\}\in \C C}
\min_{ \nu }
\left \{ 
G_u(\nu\|\mu ):\ 
\LLL \| c_n\| \le R, \ 
\{ c_n\} \in \C C_T(\nu )
\right \}
 \No
 &=&
   \inf_{ \nu }
\left \{ 
G_u(\nu\|\mu ):\ R_C(\nu )\le R
\right \} .
\Qeas
\Prend

\Tlad

{\it Proof of Theorem \ref{th:Cr*}: }
From (\ref{eq:3.35}) we have
\Eqa
\rho_C^*(R|\mu ) 
&=&
\inf_{\{ c_n\}\in \C C(R)}
\left\{
\LLL
\frac 1{\mu_n(T_n(c_n))}
:\ T_n(c_n)\ne \emptyset \mbox{ for all $n$}
\right \} .
\label{eq:4.55}
\Qea
Applying Lemma \ref{le:1} for
$B_n=T_n(c_n)\ne \emptyset$, we have
\Eq
\LLL
\frac 1{\mu_n(T_n(c_n))}
=
\min_{\nu\in \C M(\Omega )} \left\{D_u(\nu\|\mu ):
\lim_{n\to\infty} \nu_n(T_n(c_n))=1
\right\} .
\label{eq:4.56}
\Qe
It follows from (\ref{eq:4.55}), (\ref{eq:4.56}), (\ref{eq:3.25}), (\ref{eq:3.26}), and (\ref{eq:3.28}) 
that 
\Eqa
\rho_C^*(R|\mu ) 
&=&
\inf_{\{ c_n\}\in \C C(R)}
\min_{\nu}
\left\{D_u(\nu\|\mu ):
\lim_{n\to\infty} \nu_n(T_n(c_n))=1
\right \}
\No
&=&
\inf_{\{ c_n\}\in \C C}
\min_{\nu}
\left\{D_u(\nu\|\mu ):
\LLL \| c_n\| \le R, \ 
\{ c_n\} \in \C C_T(\nu)
\right \}
\No
&=&
\inf_{\nu}
\left\{D_u(\nu\|\mu ):
R_C(\nu )\le R
\right \} .
\Qea
\Prend

\Tlad

{\it Proof of Lemma \ref{le:r*kn}: }
First we will prove (\ref{eq:3.42}). 
For any $c_n\in C_n$, if $c_{0n}=c_n$ then 
\Eq
\mu_n(T_n(c_n))
\le 
\mu_n\left (\bigcup_{i=0}^{k_n}T_n(c_{0n})\right ). 
\label{eq:4.58}
\Qe
Using (\ref{eq:3.35}), (\ref{eq:3.36}) and (\ref{eq:4.58}), we obtain 
\Eq
\rho^* (R|\mu )
\ge 
\rho^* (R|\mu ; {k_n} ) .
\Qe
We now prove
\Eq
\rho^* (R|\mu )
\le 
\rho^* (R|\mu ; {k_n} )
\label{eq:4.60}
\Qe
for any non-negative integer $k_n\le n^s$. 
Let $r$ be an arbitrary real number such that $\rho^* (R|\mu ; {k_n} ) <r $. 
Then, from (\ref{eq:3.36}) and (\ref{eq:3.30}), there exists 
$\{ (c_{0n},\cdots ,c_{k_nn})\} \in \C C(R ;k_n)$
such that 
\Eq
\LLL
\frac 1{\mu_n\left (\bigcup_{i=0}^{k_n}T_n(c_{in})\right )}
<r 
\label{eq:4.61}
\Qe
and 
\Eq
\LLL \left ( \max_{0\le i\le k_n} \| c_{in} \|\right )
\le
R. 
\label{eq:4.62}
\Qe
We put 
\[
c_n^* \in \arg \max_{c} \{ \mu_n (T_n(c)):\ c=c_{in},\ i=0,\ldots ,k_n\} .
\]
Then, it is clear that  
\Eq
\| c_n^* \|
\le
\max_{0\le i\le k_n} \| c_{in} \| .
\label{eq:4.63}
\Qe
Combining (\ref{eq:4.62}) and (\ref{eq:4.63}), 
we have $\LLL \| c_n^*\|\le R$. 
Hence, from (\ref{eq:3.25}) and  (\ref{eq:3.35}), we have $\{ c_n^*\} \in \C C(R)$
and 
\Eq
\rho^*(R|\mu ) 
\le
\LLL \frac 1{\mu_n (T_n(c_n^*))} . 
\label{eq:4.64}
\Qe 
Since 
\Eqas
\mu_n\left (\bigcup_{i=0}^{k_n}T_n(c_{in})\right )
&\le&
(k_n+1) \max_{i=0,\ldots ,k_n}\mu_n (T_n(c_{in}))
\\
&=&
(k_n+1) \mu_n (T_n(c_n^*)), 
\Qeas
it follows from (\ref{eq:4.61}) and $k_n\le n^s$ that 
\Eqa
\LLL \frac 1{\mu_n (T_n(c_n^*))}
&<&
r +\LLL (k_n+1)
\No
&\le&
r. 
\label{eq:4.65}
\Qea
Combining (\ref{eq:4.64}) and (\ref{eq:4.65}), 
we have $\rho^*(R|\mu ) <r$. 
Therefore, (\ref{eq:4.60}) can be derived by setting $r= \rho^* (R|\mu ; {k_n} ) +\gamma$ ($\gamma >0$) and letting $\gamma\to 0$. 
Next, we prove (\ref{eq:3.43}). 
For any $c_n\in C_n$, if $c_{0n}=c_n$ then 
\Eq
\mu_n(T_n(c_n)^c)
\ge
\mu_n\left (\bigcap_{i=0}^{k_n}T_n(c_{in})^c\right ). 
\label{eq:4.66}
\Qe
Using (\ref{eq:3.31}), (\ref{eq:3.32}) and (\ref{eq:4.66}), we obtain
\Eq
\Ve_C (R|\mu )
\ge
\Ve_C (R|\mu ; {k_n} ) .
\Qe
We now prove
\Eq
\Ve_C (R|\mu )
\le
\Ve_C (R|\mu ; {k_n} ) 
\label{eq:4.68}
\Qe
for any non-negative integer $k_n\le n^s$. 
Let $\Ve$ be an arbitrary real number such that $\Ve_C (R|\mu ; {k_n} )<\Ve $.
Then, from (\ref{eq:3.32}) and (\ref{eq:3.30}), there exists 
$\{ (c_{0n},\cdots c_{k_nn})\} \in \C C(R;k_n)$
such that
\Eq
\LSUPn{\mu_n\left (\bigcap_{i=0}^{k_n}T_n(c_{in})^c\right )}
<
\Ve
\label{eq:4.69}
\Qe
and 
\Eq
\LLL \left ( \max_{0\le i\le k_n} \| c_{in} \|\right )
\le
R .
\label{eq:4.70}
\Qe
Since $(C_n, \|\cdot\| , T_n)$ is subadditive, for any 
$c_{0n}, c_{1n},\cdots c_{k_nn}\in C_n$, 
there exists $c_n\in C_n$ such that
\Eq
T_n(c_n)
=
\bigcup_{i=0}^{k_n}T_n(c_{in})
\label{eq:4.71}
\Qe
and 
\Eqa
\| c_n \|
&\le&
\sum_{i=0}^{k_n}\| c_{in}\|
\No
&\le&
(k_n+1)\max_{0\le i \le k_n}\| c_{in}\| .
\label{eq:4.72}
\Qea
It follows from (\ref{eq:4.70}), (\ref{eq:4.72}) and $k_n\le n^s$, 
that 
\Eqas
\LLL \| c_n \|
&\le&
R
+
\LLL (k_n+1)
\\
&\le&
R ,
\Qeas
so that 
$\{ c_n \} \in \C C(R)$. 
Hence, from (\ref{eq:3.31}), we have 
\Eq
\Ve_C (R|\mu )
\le
\LSUPn \mu_n (T_n(c_n)^c). 
\label{eq:4.73}
\Qe
Using 
(\ref{eq:4.69}) and (\ref{eq:4.71}),
we have
\Eq
\LSUPn \mu_n (T_n(c_n)^c)
<\Ve .
\label{eq:4.74}
\Qe
Combining (\ref{eq:4.73}) and (\ref{eq:4.74}), 
we have $\Ve_C (R|\mu ) <\Ve$. 
Therefore, (\ref{eq:4.68}) can be derived by setting $\Ve= \Ve_C (R|\mu ; {k_n} )+\gamma$ ($\gamma >0$) and letting $\gamma\to 0$.
Equation (\ref{eq:3.44}) can be proved in the same way as equation  (\ref{eq:3.43}). 
\Prend

\Tlad

{\it Proof of Lemma \ref{le:main0} : }
By definition of $T_n(\cdot )$, for all $c_n\in C_n$ we have 
 $$
 \C X_n^{\ c}\subset T_n(c_n)^c .
 $$ 
To simplify the proof, let $\Omega_n\setminus\C X_n \ne \emptyset$.
Then, since $\C X_n^{\ c}\ne \emptyset$, for any $c_n\in C_n$ we have 
 $$
T_n(c_n)^c\ne \emptyset .
 $$ 
Thus, from (\ref{eq:3.33}), we obtain 
\Eq
\rho_C(R|\B \mu )
=
\sup_{\{ c_n\}\in \C C(R)}
\left\{ 
\lll 
\frac 1{\mu_n(T_n(c_n)^c)}:\ 
T_n(c_n)^c\ne \emptyset
\mbox{ for all }n
\right \} .
\label{eq:4.75}
\Qe
Applying  Lemma \ref{le:1} for $B_n = T_n(c_n)^c\ne \emptyset$, we have 
\Eq
\lll
\frac 1{\mu_n(T_n(c_n)^c)}
=
\min_{\nu\in \C M(\Omega )} \left\{D_l(\nu\|\mu ):
\lim_{n\to\infty} \nu_n(T_n(c_n)^c)=1
\right\} .
\label{eq:4.76}
\Qe
Combining (\ref{eq:4.75}) and (\ref{eq:4.76}), we have 
\Eqa
\lefteqn{
\rho_C(R|\B \mu )
}\No
&=&
\sup_{c_n}
\min_{\B \nu}
\left\{
D_l(\B \nu\|\B \mu ):\ 
\lim_{n\to\infty} \nu_n(T_n(c_n)^c) =1 ,\ 
\LLL \| c_n \|\le R
\right\} .
\label{eq:4.77}
\Qea
From (\ref{eq:3.29}), for any $R$ satisfying $R_C^*(\B \nu)> R$,
we can show that  
$\LLL \| c_n \|\le R$ implies $\{ c_n\}\not\in \C C_T^*$ and 
\[
\lim_{n\to\infty} \nu_n(T_n(c_n)^c) =1 .
\]
Hence, from (\ref{eq:4.77}) we have
\Eqa
\rho_C(R|\B \mu )
&\le&
\inf_{\B \nu\in \C M(\Omega )}
\left\{
D_l(\B \nu\|\B \mu ):\ 
R^*(\B \nu)> R
\right\} .
\label{eq:4.78}
\Qea
\Prend

\Tlad

For simplicity, we put
\Eqas
\rho_C^- (R|\mu) 
&=&
\inf_{\nu\in \C M(\Omega )}\{ D_l(\nu\| \mu):\, R_C^*(\nu)\ge R\} .
\Qeas
To prove Lemma \ref{le:k2}  we need some lemmas.

\Tlad

\begin{lemma} 
\label{le:kome}
If $\rho_C^-(R|\mu )> 0$ then for all sufficiently large $n$ there exists $c_n\in C_n$ such that 
\TS
\Eq
\label{eq:4.79}
\frac 1n\log \| c_n\| \le R.
\Qe
\end{lemma}

{\it Proof: }
If 
$\rho_C^- (R|\mu)  >0$
then we obtain $R_C^*(\mu)< R$ because $D_l(\mu\|\mu ) =0$. 
Thus, by definition $R_C^*(\mu)$, there exists $\{ c_n\}\in \C C$ such that 
$$
\LLL \| c_n \| < R.
$$
Therefore, (\ref{eq:4.79})  holds for all sufficiently large $n$. 
\Prend

\Tlad

For each $a\in [0,1]$ and $\mu_n\in M(\Omega_n)$, define $R_n^a(\mu_n) $ as 
\Eq
R_n^a(\mu_n) 
=
\inf_{c\in C_n}\left\{ \frac 1n\log \| c \|:\ 
\mu_n (T_n(c)^c)\le 1-a
\right\} .
\Qe

\Tlad

\begin{lemma} 
\label{le:fin}
Let $\{ a_n\}$ be an arbitrary sequence of numbers satisfying 
\TS
\[
\lim_{n\to\infty} a_n =0 ,
\]
and let 
$\B u =\{ u_n\}, \B v =\{ v_n\}, \B w =\{ w_n\} \in \C M(\Omega)$
satisfy
$$
w_n=
\begin{cases}
u_n, 	&n\not\in I\\
v_n,	&n\in I
 \end{cases} ,
$$
where $I$ is a subset of positive integers. 
Then, 
\Eq
R_C^*(\B w)
\ge
\min\left \{ 
R_C^*(\B u ),  \liminf_{n\to\infty , n\in I} R_n^{a_n}(v_n)
 \right \} .
\label{eq:4.81}
\Qe
\end{lemma}

{\it Proof: }
Since it is obvious when $R_C^*(\B w )=\infty$, let $R_C^*(\B w )<\infty$.
Let $S$ be an arbitrary real number satisfying $S>R_C^*(\B w )$. 
Then, by the definition of $R_C^*(\cdot )$, there exists $\B c=\{ c_n\} \in \C C$ such that
\Eq
\LLL\| c_n\|<S
\label{eq:4.82}
\Qe
and
\Eq
\LINFn w_n(T_n(c_n)^c)<1 .
\label{eq:4.83}
\Qe
Equation (\ref{eq:4.83}j implies 
\Eq
\liminf_{n\to\infty, n\not\in I} u_n(T_n(c_n)^c)<1
\label{eq:4.84}
\Qe
or 
\Eq
\liminf_{n\to\infty, n\in I} v_n(T_n(c_n)^c)<1 .
\label{eq:4.85}
\Qe
In the case of (\ref{eq:4.84}),  
$$
\LINFn u_n(T_n(c_n)^c)<1
$$ 
holds, so from the definition of $R_C^*(\cdot )$,  
\Eq
\LLL\| c_n\| \ge R_C^*(\B u)
\label{eq:4.86}
\Qe
holds.
Combining (\ref{eq:4.82}) and  (\ref{eq:4.86}) we have 
$$
S> R_C^*(\B u). 
$$
In the case of (\ref{eq:4.85}), 
since there exists infinitely many $n$ such that 
$$
v_n(T_n(c_n)^c)<1-a_n, 
$$
it follows from the definition of $R_n^{a_n}(\cdot ) $ that 
\Eq
\frac 1n\log \| c_n\| \ge R_n^{a_n}(v_n). 
\label{eq:4.87}
\Qe
Combining (\ref{eq:4.82}) and (\ref{eq:4.87}) we have 
\Eqas
S
&\ge&
\liminf_{n\to\infty , n\in I} R_n^{a_n}(v_n). 
\Qeas
Consequently, we have 
\Eq
S \ge \min
\left\{ R_C^*(\B u), \liminf_{n\to\infty , n\in I} R_n^{a_n}(v_n)
\right\} .
\label{eq:4.88}
\Qe
Therefore, (\ref{eq:4.81}) can be derived by setting $S=R_C^*(\B w )+\gamma$ ($\gamma >0$) and letting $\gamma\to 0$.
\Prend

\Tlad

The following Lemma plays an important role in the proof of Lemma \ref{le:k2}.

\Tlad

\begin{lemma} 
\label{le:Em}
Let $\mu, u\in M(\Omega_n)$,  $E_m\in \C F_n$ $(m=0,1,\cdots )$
and define the measure $\nu^m $ on $\Omega_n$ 
\TS
recursively as follows: 
\begin{itemize}
\item[(i)]$\nu^0=\mu$.
\item[(ii)] If $\nu^m(E_m)\ne 0$ and $\nu^m \ne u$ then $\nu^{m+1}$ is defined by
\Eq
\label{eq:4.89}
\frac{d\nu^{m+1}}{d\nu^m}(x)=
\frac{1_{E_m}(x)}{\nu^m(E_m)}
\Qe
otherwise set $\nu^{m+1}= u$, where $1_A(x)$ denotes indicator function of set $A$.
 \end{itemize}
Put 
 $$ j = \min \{ m \ge 0 :\ \nu^m =u\} . $$
Let $l$ be an arbitrary positive integer satisfying $l\le j$. 
Put $\H E =E_0\cap E_1\cap \cdots \cap E_{l-1}$. 
Let  $\H\nu \in M(\Omega_n)$ be defined as follows: 
if $\mu (\H E) \ne 0$ 
then $\H\nu$ is defined by 
\Eq
\frac{d\H\nu}{d\mu}(x)
=
\frac{1_{\H E}(x)}
{\mu (\H E)} ,
\label{eq:4.90}
\Qe
otherwise set $\H\nu =u$. 
Then 
 \Eq
 \H\nu =\nu^l .
 \label{eq:4.91}
 \Qe
If $\H\nu\ne u$ then 
\Eq
D(\H\nu\|\mu ) 
=
-\sum_{m=0}^{l-1}\log \nu^m(E_m) .
\label{eq:4.92}
\Qe
\end{lemma}

\Tlad

The proof is given in the appendix.

\Tlad

{\it Proof of Lemma \ref{le:k2}: }
We should prove that
\Eq
\rho_C (R|\B\mu ; k_n )
\ge
\inf_{\B \nu \in \C M(\Omega )}\{ D_l(\B \nu\|\B\mu ):\ R_C^*(\B \nu )\ge R\} 
\label{eq:4.93}
\Qe
for any $k_n =[n^s]$ ($s>1$). 
Since it is obvious when $\rho_C^-(R|\mu ) = 0$,  
let $\rho_C^-(R|\mu ) > 0$.
Then, Lemma \ref{le:kome} guarantees, for sufficiently large $n$, the existence of $c_n\in C_n$ satisfying (\ref{eq:4.79}). 
Condition  (\ref{eq:3.46}) guarantees the existence of  $\B u =\{ u_n\}\in \C M(\Omega )$ satisfying 
\Eq
R_C^*(\B u )\ge R .
\label{eq:4.94}
\Qe
For simplicity, we put 
$$
a_n= \sqrt{ \frac 1{n^{s-1}}} .
$$
Then, it holds that 
$0<a_n \le 1$, 
\Eq
\lim_{n\to\infty}a_n =0, 
\label{eq:4.95}
\Qe
and 
\Eq
\lim_{n\to\infty}\frac{k_n a_n}n =\infty .
\label{eq:4.96}
\Qe
For each non-negative integer $m$, denote by $\nu^m$ the measure on $\Omega_n$ defined recursively as follows: 
\begin{itemize}
\item[(i)]
$
\nu^0=\mu_n .
$
\\
\item[(ii)] If $R_n^{a_n}(\nu^m)\ge R$ then 
$$
\nu^{m+1}=\nu^m. 
$$
If $R_n^{a_n}(\nu^m)< R$ then we choose 
$c_{mn}\in C_n$ satisfying 
\Eq
\label{eq:4.97}
\frac 1n\log \| c_{mn} \| < R
\Qe
and 
\Eq
\label{eq:4.98}
\nu^m(T_n(c_{mn})^c)\le 1- a_n, 
\Qe
and 
put
\Eq
E_m = T_n(c_{mn})^c. 
\label{eq:4.99}
\Qe
If $\nu^m \ne u_n$ and $\nu^m(E_m)\ne 0$ then $\nu^{m+1}$ is defined by
\Eq
\label{eq:4.100}
\frac{d\nu^{m+1}}{d\nu^m}(x)=
\frac{1_{E_m}(x)}{\nu^m(E_m)}
\Qe
otherwise set $\nu^{m+1}= u_n$.
\end{itemize}
Define $i_n$, $j_n$ and $l_n$ by 
\Eq
{i_n}=\min\{ m\ge 0:\ R_n^{a_n}(\nu_n^m)\ge R\}
\label{eq:4.101}
\Qe
\Eq
{j_n} =\min\{ m\ge 0:\ \nu_n^m=u_n \} ,
\label{eq:4.102}
\Qe
and 
\Eq
{l_n}=\min\{ {i_n}, {j_n} , k_n 
\} .
\label{eq:4.103}
\Qe
We put 
\Eqas
I&=& \{ n:\ l_n =i_n\}
\\
J&=& \{ n:\ l_n =j_n\}
\\
K&=& \{ n:\ l_n =k_n\} .
\Qeas
Then, clearly 
\Eq
I
=
\left \{ 
n:\ R_n^{a_n}(\nu_n^{{l_n}})\ge R
\right \}
\label{eq:4.104}
\Qe
and
\Eq
J =
\left \{ n:\ \nu_n^{{l_n}} =u_n 
\right \} .
\label{eq:4.105}
\Qe
Since ${l_n}\ne {i_n} $ implies ${l_n}={j_n}$ or ${l_n}=k_n$, we have
\Eq
I^c
\subset
J\cup (K\cap J^c\cap I^c) .
\label{eq:4.106}
\Qe
Since  $m <{l_n}$ implies $m < {i_n}$, we obtain 
\Eq
R_n^{a_n}(\nu^m)<R ,\quad m=0,1,\cdots , {l_n}-1 .
\label{eq:4.107}
\Qe
Note that  $c_{mn}$ is defined and satisfies 
\[
\frac 1n \log \| c_{mn} \| 
<
R
\]
for all $m \le {l_n}-1$.
For a subscript $m$ (${l_n}\le m \le k_n$), we define $c_{mn}$ as $c_n$ satisfying (\ref{eq:4.79}).
Then, it holds that 
\Eq
\frac 1n \log \left (\max_{m=0,1,\cdots , k_n}\| c_{mn} \| \right )
\le
R .
\label{eq:4.108} 
\Qe
If ${l_n} \ge 1$ then denote by $\H E_n $ the set defined by 
\Eq
\H E_n =
\bigcap_{m=0}^{{l_n}-1}
E_m, 
\label{eq:4.109}
\Qe
otherwise set $\H E_n =\Omega_n$. 
Since ${l_n}\le k_n$, it follows from (\ref{eq:4.99}) and (\ref{eq:4.109}) 
that 
\Eqa
\H E_n
&\supset&
\bigcap_{m=0}^{k_n}
T_n(c_{mn})^c. 
\label{eq:4.110}
\Qea
From the definition of $\rho_C (R|\B \mu ; k_n )$ and (\ref{eq:4.108}), (\ref{eq:4.110}), we obtain 
\Eq
\rho_C (R|\B \mu ; k_n
 )
\ge
\lll \frac 1{\mu_n (\H E_n)} .
\label{eq:4.111}
\Qe
Define $\H \nu_n\in M(\Omega_n)$ as follows:
If $\mu_n(\H E_n)\ne 0$ then $\H \nu_n$ is defined by 
\Eq
\label{eq:4.112}
\frac{d\H{\nu}_n}{d\mu_n}(x)
=
\frac {1_{\H E_n}(x)}{\mu_n(\H E_n)}. 
\Qe
Otherwise set $\H\nu_n= u_n$. 
Then, it is clear that $\mu_n(\H E_n)= 0$ implies
\[
-\infty =\log \mu_n (\H E_n) \le -D(\H{\nu}_n\| \mu_n ) .
\]
Thus, using (\ref{eq:4.112}) and Lemma \ref{le:1}, we have 
\Eq
\LLL \mu_n (\H E_n) \le -D_l(\H{\B{\nu}}\| \B\mu ) .
\label{eq:4.113}
\Qe
Combining (\ref{eq:4.111}) and (\ref{eq:4.113}), we have
\Eq
\rho_C (R|\B \mu ;k_n )\ge D_l(\H{\B{\nu}}\| \B\mu ) .
\label{eq:4.114}
\Qe
Since (\ref{eq:4.107}) and $l_n <j_n$ holds from (\ref{eq:4.103}), 
applying Lemma \ref{le:Em} to $l=l_n\ge 1$ we have 
\Eq
\H{\nu}_n=\nu^{{l_n}} . 
\label{eq:4.115}
\Qe
Note that (\ref{eq:4.115}) generally holds, 
since ${l_n} =0$ implies $\H{\nu}_n=\mu_n=\nu_n^0$.
Thus, (\ref{eq:4.104}) and (\ref{eq:4.105}) can be written as
\Eq
I
=
\left \{ 
n:\ R_n^{a_n}(\H\nu_n )\ge R
\right \}
\label{eq:4.116}
\Qe
and 
\Eq
J =
\left \{ n:\ \H \nu_n =u_n 
\right \} .
\label{eq:4.117}
\Qe
Put 
\Eq
K' = K\cap J^c\cap I^c .
\label{eq:4.118}
\Qe
We note here that $n\in K'$ implies $l_n=k_n$ and $\H\nu_n \ne u_n$. 
Then by using Lemma \ref{le:Em}, we have 
\Eq
D(\H\nu_n\|\mu_n ) 
=
-\sum_{m=0}^{{k_n}-1}\log \nu^m(E_m) , \mbox{ if }n\in K' .
\label{eq:4.119}
\Qe
Since $\nu^m(E_m)\le 1- a_n$ is obvious from (\ref{eq:4.98}) and (\ref{eq:4.99}), 
it follows that 
\[
D(\H\nu_n\|\mu_n)\ge -{k_n}\log \left (1-a_n\right ),
\quad \mbox{ if }n\in K' .
\]
Since 
$-\log (1-x)\ge x$ ($0<x<1$), we have 
\Eq
\frac 1n D(\H\nu_n\|\mu_n)\ge 
\frac{{k_n} a_n}n,
\quad  
\mbox{ if }n\in K' .
\label{eq:4.120}
\Qe
Thus, using (\ref{eq:4.96}), we obtain 
\Eq
\liminf_{n\to \infty , n\in K'}\frac 1n D(\H\nu_n\|\mu_n)
=
\infty .
\label{eq:4.121}
\Qe
Define $\T\nu_n \in M(\Omega_n)$ by 
\Eq
\T\nu_n=
\begin{cases}
u_n, 	&n\in K'\\
\H\nu_n,	&n\not\in K' .
 \end{cases}
\label{eq:4.122}
\Qe
Then, it follows from (\ref{eq:4.121}) and (\ref{eq:4.122}) that  
\Eq
D_l(\T{\B{\nu}}\|\B\mu )
\le
 D_l(\H{\B{\nu}}\|\B\mu ) .
\label{eq:4.123}
\Qe
Combining (\ref{eq:4.114}) and (\ref{eq:4.123}), we have 
\Eq
\rho_C (R|\B \mu ; k_n)\ge D_l(\T{\B{\nu}}\| \B\mu ). 
\label{eq:4.124}
\Qe
From (\ref{eq:4.117}) and (\ref{eq:4.122}), it can be seen that  $\T \nu_n =u_n$ holds for all $n \in J$. 
Thus, from (\ref{eq:4.122}) we obtain 
\[
J\cup K'
\subset
\left \{ n:\ \T \nu_n =u_n 
\right \} .
\]
It follows from  (\ref{eq:4.106}) and (\ref{eq:4.118}) that 
\Eq
I^c
\subset
J\cup K'
\subset 
\left \{ n:\ \T \nu_n =u_n 
\right \} .
\label{eq:4.125}
\Qe
On the other hand, since  $I\cap K' =\emptyset$ is obvious from (\ref{eq:4.118}), 
we can show that (\ref{eq:4.122})  implies 
\Eq
I
\subset
\left \{ n:\ \T \nu_n =\H\nu_n 
\right \} .
\label{eq:4.126}
\Qe
From (\ref{eq:4.125}) and (\ref{eq:4.126}) we obtain 
$$
\T\nu_n=
\begin{cases}
u_n, 	&  n\in I^c\\
\H\nu_n,& n\in I
 \end{cases} .
$$
Applying Lemma \ref{le:fin} for $w_n =\T\nu_n$, we have 
\[
R_C^*(\T \nu)
\ge
\min\left \{ 
R_C^*(\B u ),  \liminf_{n\to\infty , n\in I} R_n^{a_n}(\H\nu_n)
 \right \} .
\]
It follows from (\ref{eq:4.94}) and (\ref{eq:4.116}) that 
\Eq
R_C^*(\T{\B{\nu}})\ge R .
\label{eq:4.127}
\Qe
Equation (\ref{eq:4.93}) follows from (\ref{eq:4.124}) and (\ref{eq:4.127}). 
\Prend

\Tlad

\subsection{Proofs of results for $A_n$-code and $B_n$-code
}

We put $ C_n = A_n $ and $\C X_n = \C X_1 ^ n \times \C X_2 ^ n $, define $ \| \cdot \| $ by (\ref {eq:2.1}), and define $T_n(\cdot )$ by (\ref{eq:2.2}). 
Let $\mu_n$ be a probability measure of $(X_n^1, X_n^2)$. 
Then, it holds that $\Ve_A(R) =\Ve_A(R|\mu )$, $\rho_A(R)=\rho_A(R|\mu )$, $\rho_A^*(R)=\rho_A^*(R|\mu )$, and so on.
Similarly, $\Ve_B(R) =\Ve_B(R|\mu )$, $\rho_B(R)=\rho_B(R|\mu )$, and so on.
It is notice that $\mu_n \in M(\C X_n )\subset M(\Omega_n)$. 
By Theorem \ref{th:MK}, for any $\nu\in \C M(\Omega )$ we have  
\Eq
R_A(\nu )=R_B(\nu ), \quad
R_A^*(\nu )=R_B^*(\nu ). 
\label{eq:4.128}
\Qe

Now we show the subadditivity of the $A_n$-code.

\Tlad

\begin{prop}
\label{pr:an}
$(A_n, \| \cdot \| , T_n)$ is subadditive.
\TS
\end{prop}

{\it Proof: } 
For any $c_1=(\varphi_1, \psi_1)$ and $c_2=(\varphi_2, \psi_2)\in A_n$ 
we define $c=(\varphi ,\psi )\in A_n$ with 
$$
\varphi (x,y)=
\begin{cases}
\varphi_1(x,y), 				&(x,y) \in T_n(c_1 )\\
\varphi_2(x,y)+\| c_1\|,	&\mbox{otherwise}
 \end{cases} 
$$
and 
$$
\psi (k,y) =
\begin{cases}
\psi_1(k,y), 				&k=1,2,\cdots , \| c_1\|\\
\psi_2(k-\| c_1\| ,y),	&\mbox{otherwise}.
 \end{cases}. 
$$
Then, noting the definition of 
$(A_n, \|\cdot \| ,T_n)$ implies that 
$c(x,y)=\psi (\varphi(x,y),y)$, which yields
$$
c(x,y)=
\begin{cases}
c_1(x,y), 				&(x,y)\in T_n(c_1)\\
c_2(x,y),	&\mbox{otherwise}
 \end{cases} .
$$
Furthermore, by noting $T_n(c)=\{ (x,y):\ c(x,y)=x\}$, it follows that 
$T_n(c)=T_n(c_1)\cup T_n(c_2) $. 
Since $\| c\|$  is the number of elements of the domain of the one-variable function $\psi (k , y)$ with $k$ as variable,  $\|c \| = \|c_1\|+\|c_2\|$ holds.
This indicates that $(A_n, \|\cdot \| , T_n)$ is subadditive. 
\Prend

\Tlad

\Tlad

{ \it Proof of Theorem \ref{th:ABe}: }
Using Theorem \ref{th:Ce} and  (\ref{eq:4.128}), we have 
\Eqa
\frac 1{1-\Ve_A(R|\mu )}
&=&
\inf_{\nu :\ R_A(\nu )\le R} G_u(\nu\|\mu )
\No
&=&
\inf_{\nu :\ R_B(\nu )\le R} G_u(\nu\|\mu )
\No
&=&
\frac 1{1-\Ve_B(R|\mu )} .
\label{eq:4.129}
\Qea
By definition, clearly 
\Eq
\Ve_B(R|\mu )
\ge
\Ve_B(R|\mu ; k_n). 
\label{eq:4.130}
\Qe
Since $B_n\subset A_n$, by the definition of $\Ve_C(R|\mu ; k_n)$, we have
\Eq
\Ve_B(R|\mu ; k_n) 
\ge
\Ve_A(R|\mu ; k_n) .
\label{eq:4.131}
\Qe
Since $(A_n, \| \cdot \|, T_n )$ is subadditive, from Lemma \ref{le:r*kn} we obtain 
\Eq
\Ve_A(R|\mu ) =\Ve_A(R|\mu ; k_n) 
\label{eq:4.132}
\Qe
for any non-negative integer $k_n\le n^s$. 
Combining (\ref{eq:4.129})--(\ref{eq:4.132}), we have
 $\Ve_A(R|\mu )=\Ve_B(R|\mu ) =\Ve_B(R|\mu ; k_n)=\Ve_A(R|\mu ; k_n) $. 
\Prend

\Tlad

{\it Proof of Theorem \ref{th:ABr}: }
Without loss of generality, we may assume that $\Omega_n\setminus \C X_n\ne \emptyset$.
Then, note that the condition (\ref{eq:3.46}) holds. 
Since $(A_n, \| \cdot \|, T_n )$ is subadditive, the combination of Lemma \ref{le:r*kn} and Theorem \ref{th:Cr} leads to 
\Eq
\rho_A(R|\mu) =\rho_A(R| \mu ;{k_n} ) =\rho_A^- (R|\mu) 
\label{eq:4.133}
\Qe
for all continuous point $R$ of $\rho_A(R|\mu)$, where $k_n=[n^s]$ ($n=1,2,\cdots$). 
Using (\ref{eq:4.128}) and the definition of $\rho_C^- (R|\mu)$, we obtain 
\Eq
\rho_A^- (R|\mu) =\rho_B^- (R|\mu) . 
\label{eq:4.134}
\Qe
Applying Lemma  \ref{le:k2}  to $(B_n, \| \cdot \|, T_n )$ yields
\Eq
\rho_B(R| \mu ;{k_n} ) \ge \rho_B^- (R|\mu) .
\label{eq:4.135}
\Qe
Combining  (\ref{eq:4.133})--(\ref{eq:4.135}), we have
\Eq
\rho_A(R| \mu )=\rho_A(R| \mu ;{k_n} ) . 
\le
\rho_B(R| \mu ;{k_n} ) 
\label{eq:4.136}
\Qe
On the other hand, since $B_n\subset A_n$, by the definition of $\rho_C (R|\mu ;{k_n})$ we have 
\Eq
\rho_A(R| \mu ;{k_n} ) 
\ge
\rho_B(R| \mu ;{k_n} ) .
\label{eq:4.137}
\Qe
Combining (\ref{eq:4.136}) and (\ref{eq:4.137}), we obtain 
$$
\rho_A(R| \mu ) 
=
\rho_A(R| \mu ;{k_n} ) 
=
\rho_B(R| \mu ;{k_n} ) .
$$
\Prend

\Tlad

{\it Proof of Theorem \ref{th:ABr*}: }
Using Theorem \ref{th:Cr*}, Lemma \ref{le:r*kn} and  (\ref{eq:4.128}), we have 
\Eqas
\rho_A^*(R|\mu ) 
&=&
\inf_{\nu\in \C M(\Omega )}\{ D_u(\nu\| \mu):\, R_A(\nu)\le R\}
\\
&=&
\inf_{\nu\in \C M(\Omega )}\{ D_u(\nu\| \mu):\, R_B(\nu)\le R\}
\\
&=&
\rho_B^*(R|\mu ) 
\\
&=&
\rho_B^*(R|\mu ; k_n) .
\Qeas
\Prend

\section{
Concluding remarks}
\label{se:con}

In this paper, based on the concept of "representing the performance of one coding system $A_n$ by the number of other coding systems $B_n$", we considered the merits of an encoder that can access side information.
Remark \ref{re:ABe} states that research under this concept has not only theoretical interest but also practical value. 
In general, if there is one device $A_n$ with the same performance as many devices $B_n$, it would be better to replace many devices with one device. 
In order to determine the number $k_n$ of devices that can be reduced, it is important to determine the number of other devices $B_n$ having the same performance as one device $A_n$.
Theorems \ref{th:ABe}, \ref{th:ABr} and \ref{th:ABr*} have given the number $k_n$ of $B_n$ that can be reduced by changing the coding system from $B_n$ to $A_n$.
Whether this concept can be applied to study other multi-terminal coding problems is a future subject.

In this paper we have obtained general expressions for the upper limit of the error probability and the reliability functions (Theorem \ref{th:Ce}, \ref{th:Cr*} and \ref{th:Cr}). 
Theorems \ref{th:Ce} and \ref{th:Cr*} hold unconditionally.
On the other hand, Theorem \ref{th:Cr} gives a sufficient condition to express 
\Eq
\rho_C (R|\mu )
=
\inf_{\nu:\, R_C^*(\nu)> R } D_l(\nu\| \mu) .
\label{eq:5.138}
\Qe
When applied to source coding, it is not unnatural to assume the condition (\ref{eq:3.46}). 
For example, in the case of fixed-length lossless source coding, if the source alphabet is a countable set, there exists a sequence $\nu =\{ \nu_n\}$  of uniform distribution $\nu_n(x)$ satisfying $R_C^*(\nu)\ge R$.
Thus, by assuming that $(C_n, \| \cdot \| , T_n)$ is subadditive, the reliability function $\rho_C (R|\mu )$ can be expressed as (\ref{eq:5.138}) for any continuous points $ R $ of $\rho_C (\cdot |\mu )$. 
Therefore, it is an important issue to clarify the conditions under which subadditivity holds.
Proposition \ref{pr:an} means that fixed-length lossless source coding is subadditive if the side information is accessible for both encoding and decoding.
It is easy to generalize this result to the lossy case. 
Assuming that the distortion function is $d_n(\cdot ,\cdot )$ and $\H T_n(c)$ is defined by  
$$
\H T_n( c)=\{ (x,y)\in \C X_1^n\times\C X_2^n:\ 
d_n(x, c(x,y))\le D
\}
$$ for each  $c\in A_n$, it can be easily shown that  $(A_n, \| \cdot \| , \H T_n)$ is subadditive. 
That is, even in the lossy case, (\ref{eq:5.138}) holds for continuous points $ R $ of $\rho_C (\cdot |\mu )$, where $C_n$ is replaced with $A_n$.
Then it should be noted that $\Ve_n(c_n) =\mu_n (\H T_n( c_n)^c)$ is the probability that distortion is greater than $D$, and $R_A^*(\nu)$ is a rate-distortion function.
The fact that formula (\ref{eq:5.138}) can be derived not only in the lossless case but also the lossy case will justify the usefulness of Theorem \ref{th:Cr} and the concept of subadditivity.
As an application of Theorem \ref{th:Cr}, we can derive the results of \cite{Ir01,IrIh} on the reliability function in fixed-length lossless and lossy source coding.
Since Theorem \ref{th:Cr} holds without using an information-theoretic specific structure, we are convinced that it is an important theorem applicable to various fields.
Indeed, applying Theorem \ref{th:Cr} to hypothesis testing yields an expression for the supremum $r$-achivable error probability exponents $B_e(r|\X\| \O{\X})$ defined in \cite{H00b,H03}. 
The expression is similar to, but not identical to, that in Theorem 2 of \cite {Ir05}.
Details are omitted here.

The results of $(C_n, \| \cdot \| , T_n)$ can also be applied to the study of multi-terminal coding systems, since no specific information theoretical structure is assumed.
In particular, Lemma \ref{le:r*kn} suggests that the concept of subadditive can play an important role in the study of list decoding. 
Recently, list decoding has been studied not only in channel coding problems, but also in other problems such as source coding problems and biometric identification problems (cf. \cite{SUM,YY01}).
It is a future subject to clarify the relationship between these problems and subadditivity.


\setcounter{section}{0}
\app{{\normalsize P}ROOF OF {\normalsize L}EMMA {\normalsize  \ref{le:lem2}}}
To prove Lemma \ref{le:lem2} we need a lemma. 

\Tlad

\begin{lemma} 
\label{le:lem1}
For any $\mu \in M(\Omega_n )$ and any $B\in \C F_n$ ($B\ne \emptyset$), 
\TS
\Eq
\frac 1{\mu (B)} 
=
\min_{\nu \in M(B)}G(\nu\|\mu ) .
\label{eq:A.1}
\Qe
\end{lemma}

{\it Proof: }
If $\mu (B)=0$, then 
the both sides of 
(\ref{eq:A.1})
are equal to $\infty$, because 
$G(\nu ||\mu )=\infty$ 
 for each $\nu \in M(B)$. 
We now assume that $\mu (B)>0$. 
Define a measure $\mu^* \in M(B)$ by 
\[
\frac{d\mu^*}{d\mu}(x)
=
\frac{1_B(x)}{\mu (B)},\quad  x\in \Omega_n.
\]
Then for any $\nu \in M(B)$ satisfying $G (\nu ||\mu )<\infty$, 
we have 
\Eqa
 G (\nu ||\mu )
&\ge&
\int_B\frac{d\nu}{d\mu}(x)d\nu (x)
\No
&=&
\int_B\frac{d\mu^*}{d\mu}(x)\frac{d\nu}{d\mu^*}(x)d\nu (x)
\No
&=&
\frac 1{\mu (B)}
\int_B\frac{d\nu}{d\mu^*}(x)d\nu (x)
\No
&=&
\frac 1{\mu (B)}
\int_B \left (\frac{d\nu}{d\mu^*}(x) \right )^2d\mu^* (x)
\No
&\ge&
\frac 1{\mu (B)}
\left (\int_B\frac{d\nu}{d\mu^*}(x)d\mu^* (x)\right )^2
\No
&=&
\frac {\nu (B)^2}{\mu (B)}
\No
&=&
\frac 1{\mu (B)}
\label{eq:A.2}
\Qea
Noting that 
(\ref{eq:A.2}) 
holds with 
equality for $\nu = \mu^*$, 
we obtain 
(\ref{eq:A.1}) 
from 
(\ref{eq:A.2}).  
\Prend

\Tlad

{\it Proof of Lemma \ref{le:lem2}: }
Define 
$\mu^* = \{ \mu_n^* \} \in \C{M}(\B{B})$ as follows: 
if
$\mu_n(B_n)\ne 0$ 
then $\mu^*_n$ is defind by 
\Eq
\frac{d\mu^*_n}{d\mu_n}(x)
=
\frac{1_{B_n}(x)}{\mu_n(B_n)},\quad 
 x \in \Omega_n ,
\label{eq:A.3}
\Qe
otherwise let $\mu_n^*$ be an arbitrary fixed element of $M(B_n)$. 
Then, as is seen in the proof of 
Lemma 
\ref{le:lem1},  
\[
\frac 1{\mu_n(B_n)}= G (\mu^*_n||\mu_n)
\]
and we have 
\Eqa
\LINFn \frac 1{\mu_n(B_n)}
&=&
G_l (\mu^*||\mu )
\label{eq:A.4}
\\
&\geq&
\inf_{ \nu \in \C{M}(\B{B})} 
G_l(\nu ||\mu ).
\label{eq:A.5}
\Qea
Let $\nu = \{ \nu_n \} \in \C{M}(\B{B})$ 
be an arbitrary sequence. 
We define 
$\nu^* = \{ \nu_n^* \} \in \C{M}(\B{B})$ by 
\[
\frac{d\nu^*_n}{d\nu_n}(x)
=
\frac{1_{B_n}(x)}{\nu_n(B_n)},\quad x \in \Omega_n.
\]
Then, since 
\Eqas
 G (\nu_n^*||\mu_n)
&=&
\frac 1{\nu_n(B_n)^2}
\int_{B_n}\frac{d\nu_n}{d\mu_n}(x)d\nu_n(x),
\Qeas
we have 
\Eqas
G_l(\nu^*||\mu )
&=&
\LINFn
\int_{B_n}  \frac{d\nu_n}{d\mu_n}(x)d\nu_n(x).
\No
&\le&
G_l(\nu ||\mu )
\Qeas
Since $\nu^*_n(B_n)=1$, using 
Lemma 
\ref{le:lem1}, 
we have 
\[
\frac 1{\mu_n(B_n)}
=
\min_{\nu_n\in M(B_n)}G (\nu_n||\mu_n)
\le
G (\nu^*_n||\mu_n ).
\]
Therefore 
\[
\LINFn \frac 1{\mu_n(B_n)}
\le
G_l(\nu^*||\mu )
\le
G_l(\nu ||\mu )
\]
and 
\Eq
\LINFn \frac 1{\mu_n(B_n)}
\le
\inf_{ \nu \in \C{M}(\B{B})} 
G_l(\nu ||\mu ).
\label{eq:A.6}
\Qe
Equation 
(\ref{eq:4.51}) 
follows from 
(\ref{eq:A.4}), 
(\ref{eq:A.5})
and 
(\ref{eq:A.6}). 
Equation 
(\ref{eq:4.52}) 
can be proved similarly. 
\Prend

\Tlad

\app{{\normalsize P}ROOF OF {\normalsize L}EMMA {\normalsize   \ref{le:Em} }}
To prove Lemma \ref{le:Em} we need a lemma. 

\Tlad

\begin{lemma} 
\label{le:2}
Let $l$ be an arbitrary positive integer. 
For any $\nu^0, \nu^{m+1}\in M(\Omega_n)$ and $E_m\in \C F_n$
\TS
 ($m=0,1,\cdots ,l-1$)
satisfying (\ref{eq:4.89}), it holds that 
\Eq
\frac{d\nu^{l}}{d\nu^0}(x)=
\frac{1_{E_0\cap E_1\cap\cdots \cap E_{l-1}}(x)}
{\nu^0 (E_0\cap E_1\cap\cdots \cap E_{l-1})} ,
\label{eq:B.1}
\Qe
and 
\Eq
D(\nu^l\| \nu^0)
=
-\sum_{m=0}^{l-1}
\log \nu^m(E_m).
\label{eq:B.2}
\Qe
\end{lemma}

{\it Proof: }
From (\ref{eq:4.89}) we have 
$$
\nu^{1}(E_1)
=
\frac{\nu^0(E_0\cap E_1)}{\nu^0(E_0)}
$$
and 
$$
\nu^{2}(E_2)
=
\frac{\nu^1(E_1\cap E_2)}{\nu^1(E_1)}
=
\frac{\nu^0(E_0\cap E_1\cap E_2)}{\nu^0(E_0)\nu^1(E_1)}. 
$$
Similarly it holds that
$$
\nu^{l-1}(E_{l-1})
=
\frac{\nu^0(E_0\cap E_1\cap \cdots \cap E_{l-1})}{
\nu^0(E_0)\nu^1(E_1)\cdots \nu^{l-2}(E_{l-2})} ,
$$
so that 
\Eq
\nu^0 (E_0\cap E_1\cap\cdots \cap E_{l-1})
=
\prod_{m=0}^{l-1}\nu^m(E_m) .
\label{eq:B.3}
\Qe
Noting $1_{A\cap B}(x) =1_A(x)1_B(x)$ we obtain  
\Eq
1_{E_0\cap E_1\cap\cdots \cap E_{l-1}}(x)
=
\prod_{m=0}^{l-1}1_{E_m}(x) .
\label{eq:B.4}
\Qe
Using (\ref{eq:4.89}) we have 
\Eqa
\frac{d\nu^{l}}{d\nu^0}(x)
&=&
\prod_{m=0}^{l-1}
\frac{d\nu^{m+1}}{d\nu^m}(x)
\No
&=&
\prod_{m=0}^{l-1}
\frac{1_{E_m}(x)}{\nu^m(E_m)} .
\label{eq:B.5}
\Qea
Combining (\ref{eq:B.3})--(\ref{eq:B.5})
we have (\ref{eq:B.1}). 
From (\ref{eq:B.5})
we have 
\Eqas
D(\nu^l\| \nu^0)
&=&
\int_{\Omega_n}\log \left (\prod_{m=0}^{l-1}
\frac{1_{E_m}(x)}{\nu^m(E_m)} \right )d\nu^l(x)
\\
&=&
\sum_{m=0}^{l-1}
\int_{\Omega_n} \log 
\frac{1_{E_m}(x)}{\nu^m(E_m)} d\nu^l(x) .
\Qeas
Since $\nu^l(E_m)=1$ holds for all $m\le l-1$, 
we have (\ref{eq:B.2}). 
\Prend

\Tlad

{\it Proof of Lemma \ref{le:Em}: }
Let $l$ be an arbitrary positive integer satisfying $l\le j$. 
First consider the case of $\nu^l\ne u$. 
In this case, by definition,  $\nu^m\ne u$ holds for all $m\le l$. 
This means that (\ref{eq:4.89}) holds for all $m=0,1,\cdots ,l-1$. 
Thus, by Lemma \ref{le:2}, we have 
\Eq
D(\nu^l\|\mu ) 
=
-\sum_{m=0}^{l-1}\log \nu^m(E_m) , \quad \nu^l\ne u
\label{eq:B.6}
\Qe
and 
\Eq
\frac{d\nu^l}{d\mu}(x)=
\frac{1_{E_0\cap E_1\cap\cdots \cap E_{l-1}}(x)}
{\mu (E_0\cap E_1\cap\cdots \cap E_{l-1})}. 
\label{eq:B.7}
\Qe
From (\ref{eq:4.90}) and (\ref{eq:B.7}) we have $\H\nu =\nu^l$.
Next, consider the case of $\nu^l =u$. 
By the definition of $j$ we have that $l\le j$ implies $\nu^{l-1}\ne u$, which implies (\ref{eq:4.89}) holds for all $m\le l-2$. 
If 
$$
\nu^{l-1}(E_{l-1}) \ne 0
$$ 
then (\ref{eq:4.89}) also holds for $m=l-1$. 
Thus, by Lemma \ref{le:2} we have (\ref{eq:B.7}), so $\H\nu =\nu^l$.
We now consider the case where 
\Eq
\nu^{l-1}(E_{l-1}) =0. 
\label{eq:B.8}
\Qe
Since $\nu^{l-1}\ne u$, replacing $l$ with $l-1$ in (\ref{eq:B.7})  we have
\Eq
\frac{d\nu^{l-1}}{d\mu}(x)=
\frac{1_{E_0\cap E_1\cap\cdots \cap E_{l-2}}(x)}
{\mu (E_0\cap E_1\cap\cdots \cap E_{l-2})}
\label{eq:B.9}
\Qe
Accordingly, 
\Eq
\nu^{l-1}(E_{l-1})
=
\frac{\mu (E_0\cap E_1\cap\cdots \cap E_{l-2} \cap E_{l-1})}
{\mu (E_0\cap E_1\cap\cdots \cap E_{l-2})} .
 \label{eq:B.10}
\Qe
From (\ref{eq:B.8}) and (\ref{eq:B.10}) we have 
$$
\quad \mu (E_0\cap E_1\cap\cdots \cap E_{l-1})
=
0. 
$$
Thus, by the definition of $\H\nu$, we obtain $\H\nu = u=\nu^l $. 
Consequently, (\ref{eq:4.91}) holds. 
Equation (\ref{eq:4.92}) follows from (\ref{eq:B.6}).  
\Prend

\Tlad

\providecommand{\bysame}{\leavevmode\hbox to3em{\hrulefill}\thinspace}

\end{document}